\documentclass[%
 reprint,
 onecolumn,
 amsmath,amssymb,
 aps,
floatfix,
]{revtex4-2}

\usepackage{xcolor}
\usepackage{comment}
\usepackage{graphicx}
\usepackage{dcolumn}
\usepackage{bm}

\begin{document}

\preprint{APS/123-QED}

\title{Cavity dynamics after the injection of a microfluidic jet in capillary bridges}

\author{Miguel A. Quetzeri-Santiago$^*$ and David Fernandez Rivas}
 \email{m.a.quetzerisantiago@utwente.nl}
 \email{d.fernandezrivas@utwente.nl}
\affiliation{%
 Mesoscale Chemical Systems Group, MESA+ Institute and Faculty of Science and Technology, University of Twente, P.O. Box 217, 7500AE Enschede, The Netherlands}%
\date{\today}

\begin{abstract}
The impact of solid and liquid objects (projectiles) onto liquids and soft solids (targets) generally results on the creation and expansion of an air cavity inside the impacted objects. The dynamics of cavity expansion and collapse depends on the projectile inertia as well as on the target properties. In this paper we study the impact of microfluidic jets generated by thermocavitation processes on a capillary bridge between two parallel planar walls. Different capillary bridge types were studied, Newtonian liquids, viscoelastic liquids and agarose gels. Thus, we compare the cavity formation and collapse between a wide range of material properties. Moreover, we model the critical impact velocity for a jet to traverse a capillary bridge type. Our results show that agarose gels with a storage modulus lower that 176 Pa can be modelled as a liquid for this transition. However, the predicted transition deviates for agarose gels with higher storage modulus. Additionally, we show different types of cavity collapse, depending on the Weber number and the capillary bridge properties. We conclude that the type of collapse determines the number and size of entrained bubbles. Furthermore, we study the effects of wettability on the adhesion forces and contact line dissipation. We conclude that upon cavity collapse, for hydrophobic walls a Worthington jet is energetically favourable. In contrast, for hydrophilic walls, the contact line dissipation is in the same order of magnitude of the energy of the impacted jet, suppressing the Worthington jet formation. Our results provide strategies for preventing bubble entrapment and give an estimation of the cavity dynamics for needle-free injection applications and additive manufacturing among other applications.  
\end{abstract}


\maketitle


\section{\label{sec:level1}Introduction}

The impact of droplets and jets onto other liquids and solid objects is a recurrent phenomenon, both in nature, industrial and medical contexts \cite{prosperetti1993impact,rein1993phenomena,josserand2016drop}. Rain droplets impact bodies of water, leaves or the soil \cite{prosperetti1993impact,gilet2015fluid, sun2022stress}; droplets for inkjet printing and additive manufacturing impact previously deposited liquid layers or dry paper \cite{castrejon2013future, quetzeri2019additive}; and in needle-free injections a liquid jet is directed to impact and penetrate the skin \cite{schoppink2022jet}.
Droplet impacts onto pools have been studied since the works of Arthur Mason Worthington at the start of the 20th century \cite{worthington1900iv}. With the development of high speed cameras more features of these phenomena have been discovered and disentangled. However, the input parameters, the outcomes, and practical applications are so diverse that is still an active research topic \cite{truscott2014water, speirs2018entry, deka2018dynamics, eshraghi2020seal}. 

Previous research on water entry of liquid and spheres has focused on the critical energy necessary for air entrainment into the pool, the collapse of the entrained air cavity and the subsequent formation of a liquid jet that travels in the opposite direction of impact, i.e., a Worthington jet \cite{oguz1995air, lee1997cavity, zhu2000mechanism}. The time of collapse, cavity geometry and Worthington jet, depend on both, the properties of the liquid pool and the impacting object \cite{truscott2014water}. The phenomena is usually well described by the ratio between the surface tension, inertia and hydrostatics. Regime maps are often constructed in terms of the Froude number $Fr = U_0^2/(gD_0)$ and the Weber number $We = \rho_0 D_0 U_0^2 / \gamma_{cb}$, where $\rho_0$ and $D_0$ are the density and diameter of the impacting object, $U_0$ is the impact velocity, $\gamma_{cb}$ is the surface tension of the impacted object and g is the acceleration due to gravity \cite{aristoff2009water}. Recently, research has been extended to impact on non-Newtonian liquids as well as soft solids, such as hydrogels \cite{de2019high,kiyama2019gelatine}. Results on hydrogels show resemblance of the cavities, shape, and closure to those formed in water, albeit, for hydrogels, elasticity needs to be considered. An effective parameter to introduce the elastic properties on the collapse of the cavity is the elastic Froude number $Fr_e = \rho U_0^2 /G$, where G is the shear modulus\cite{kiyama2019gelatine}. 

Past works mainly focus on the impact onto semi-infinite pools and proximity effects of solid interfaces remain relatively unexplored \cite{zou2012large, zou2013phenomena, guo2020investigation}. Zou et al. in 2013, studied the impact of liquid droplets on pools contained in tubes with different diameters \cite{zou2013phenomena}. The study concluded that the cavity collapse and entrainment of a large bubble was dependent on the Weber number and the distance from the surrounding walls to the impact point \cite{zou2013phenomena}.  Furthermore, the impact of micrometer sized projectile at $Fr >> 1$ and $We >> 1$, where the cavity collapse is driven by the capillary forces instead of the hydrostatic pressure, has not been thoroughly described. This impact regime is relevant in processes such as 3D printing, spray painting and needle-free injections, as the relevant length scale is in micrometers and gravity does not influence such processes \cite{antkowiak2011instant,herczynski2011painting, klein2015drop,schoppink2022jet}. 

Here, we study the impact of microfluidic jets onto millimetre sized droplets and agarose gels confined between two glass slides, i.e., capillary bridges.  From a perpendicular view of the impact, we can compare the cavity formation and collapse between a wide range of material properties. We study the formation and collapse of the air cavity upon impact and extract the cavity profiles for each frame. Furthermore, by changing the wettability of the walls we study the influence of the liquid-gas interface curvature and the contact line forces on the impact outcome. These experiments give us valuable information of jet interaction with materials ranging from liquids (Newtonian and viscoelastic) to soft solids (agarose gels). The impact conditions we describe in this study, namely the size and impact speed of the jets, are similar to those of practical relevance, including needle-free injection systems and additive manufacturing. Thus, our results advance the understanding of such applications. 


\section{Experimental methods}

\begin{figure}
\centering
\includegraphics[width=15cm]{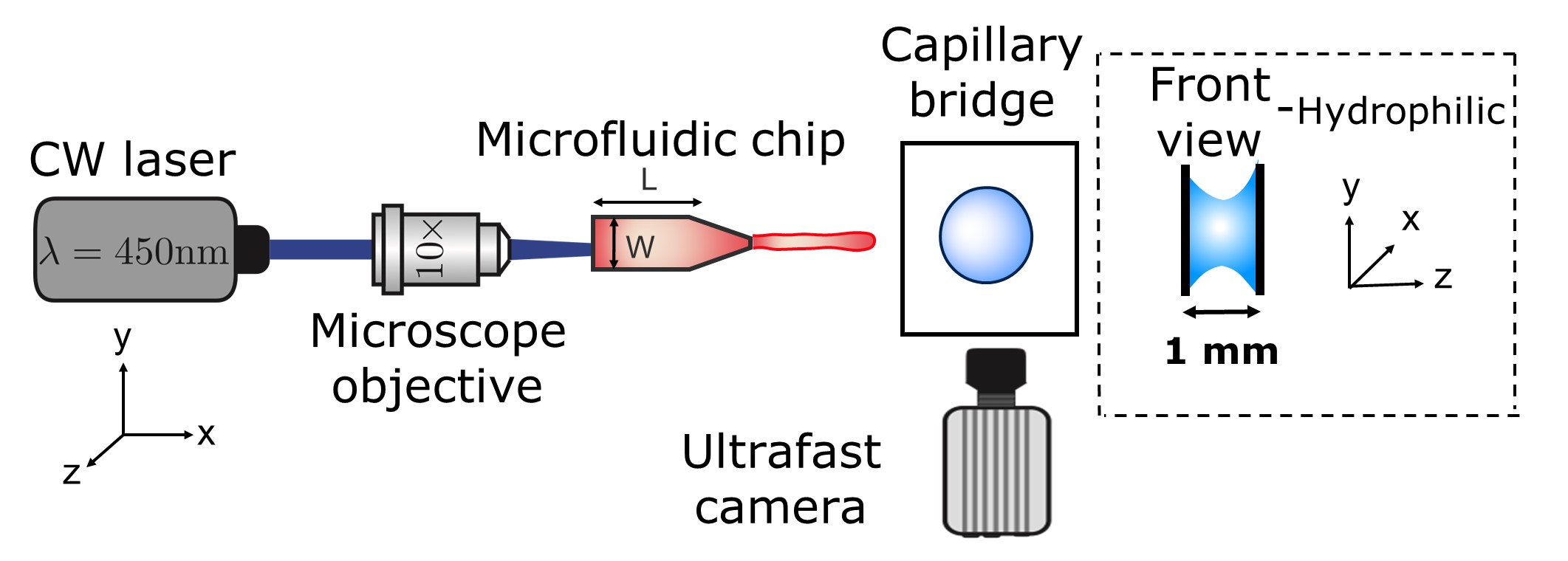}
\caption{Schematic of the experimental setup. A microfluidic jet is generatated by a thermocavitation process and is directed to impact a liquid on droplet or agarose droplet placed in between two glass slides separated by 1 mm, forming a capillary bridge. The process is recorded via shadowgraph imaging with a high-speed camera.}
\label{Fig1}
\end{figure}

To create capillary bridges, liquid or agarose gel (before curing) droplets were confined between two Borofloat glass slides standing apart 1 mm. The droplets were generated with an Eppendorf pipette and placed in between the opening of the glass slides. In this way the droplet was driven by capillary forces until it adopted a circular shape as described in ref. \cite{bouchard2019droplet}. The liquids used partially wet glass with a contact angle of 23 degrees. Capillary forces maintained the droplet pinned once equilibrium was reached. Experiments were also conducted in capillary bridges between hydrophobic walls. In this case, the glass walls where coated with Glaco spray coating. For the latter, the coated glass walls were positioned parallel to the xz plane (rather than parallel to the xy plane), as the capillary bridge would slide otherwise. The camera was positioned parallel to the y axis.   

The  capillary bridges were formed with three distinct type of materials; Newtonian liquids with different viscosities, viscoelastic solutions with different relaxation times and agarose gels with different storage modulii.  The Newtonian liquids used were water and an aqueous glycerol solution at 78 wt$\%$. The viscoelastic liquids were water based polyethylene-oxide (PEO) solutions of molecular weight ranging from 600 kDa to 1000 kDa at different concentrations. Agarose gels, were prepared by diluting agarose powder (OmniPur agarose, CAS No. 9012-36-6) in Milli-Q water at different concentrations and heated up 45 s in a microwave at 700 W. The solutions were cooled down until they reached 50 degrees Celsius before casting between the glass slides. All chemical compounds were acquired from Sigma-Aldrich. 

We impacted the capillary bridges with microfluidic jets generated through thermocavitation with velocities, and diameters in the range of $U_0 = [8 - 69.5]$ m/s and $D_0 = [50-120]$ $\mu$m, respectively. Thermocavitation is a process where liquid is vaporised and generates an expanding bubble \citep{rastopov1991sound, padilla2014optic}. This expanding bubble subsequently pushes the liquid in front of it, creating a liquid jet \citep{rodriguez2017toward, oyarte2020microfluidics}. 

The experimental setup, as shown in figure \ref{Fig1}, consist of a Borofloat glass microfluidic chip, which acts as a reservoir for the liquid and controls the jet ballistics \citep{oyarte2020microfluidics}. The liquid used for the jet is a water solution containing a red dye (Direct Red 81, CAS No. 2610-11-9) at 0.5 wt.$\%$. The red dye enhances the  laser energy absorption from a continuous wave laser $\lambda = 450$ nm (Roithner LaserTechnik, nominal power of 3.5 W). The laser is focused at the microfluidic chip with a 10x objective. The liquid jet impacting speed $U_{0}$ and its diameter $D_{0}$ were controlled with the microchannel geometry, its distance to the focal point of the microscope objective and by varying the laser power from 0.4 W to 2.1 W. A detailed description of the system can be found elsewhere \cite{quetzeri2021impact}.

The surface tension of all the liquids was measured with the Pendent Drop ImageJ plugin \citep{daerr2016pendent_drop}. The surface tension $\gamma$ and shear viscosity $\mu$ of the agarose gels was obtained through the following linear relations \cite{shao2020method},

\begin{equation}
    \gamma = 0.001022\ G + 0.07229
    \label{SurfaceTensionAgarose}
\end{equation}

\begin{equation}
    \mu = 6.005 \times 10^{-5}G + 0.00834.
\end{equation}

The rheological properties of all the materials were measured with an Anton Paar MCR 502 rheometer. The properties of the liquids and agarose gels are reported in table \ref{FluidProperties}. The shear viscosity ($\mu$) of the different capillary bridges spans two orders of magnitude, while their surface tension ($\gamma$) spans three orders of magnitude. 

The processes of bubble generation, jet ejection and impact on the capillary bridges were recorded with a Photron Fastcam SAX coupled with a 2x microscope objective. A typical experiment duration was $\sim 5$ ms and the camera resolution was set to $768 \times 328$ pixels$^2$ at a sample rate of $50$k frames per second with an exposure time of $2.5$ $\mu$s. Image analysis to extract the jet diameter, impact velocity and cavity dynamics was performed with a custom generated MATLAB script.

\begin{table}
\small
  \caption{List of fluids and agarose  capillary bridge used providing their shear viscosity $\mu$, surface tension $\gamma$ and density $\rho$. The viscoelastic relaxation time $\lambda$ is also shown for the polyethylene-oxide solutions and the storage modulus G for the agarose gels}
  \label{FluidProperties}
  \begin{center}
  \begin{tabular*}{\textwidth}{@{\extracolsep{\fill}}lrrrcc}
    \hline
   Fluid/Gel & $\mu$ (mPa s) & $\gamma$ (mN/m) & $\rho$ (kg/m$^3$) &  $\lambda$ (ms) & G(Pa)\\
    \hline
    Water & 1.0 & 72.1 & \ \ 998 & - & -\\
    Glycerol 78 wt.$\%$ & 43.60 & 65.2 & 1212 & - & -\\
    Water $\&$ red dye 0.5 wt.$\%$ & 0.91 & 47.0 & 1000 & - & -\\
    PEO 600k 0.1 wt.$\%$ & 1.56 & 63.1 & 996 & 0.31 $\pm$ 0.04 & -\\
    PEO 600k 1.0 wt.$\%$ & 21.70 & 62.9 & 998 &  1.32 $\pm$ 0.08&-\\
    PEO 1M 1.0 wt.$\%$& 44.74 & 59.2 & 998 & 6.14 $\pm$ 0.69&-\\
    Agarose 0.15 wt.$\%$ & 18.96 & 252.9 & 1000& -  & 176 \ $\pm$ \ 29  \\
    Agarose 0.25 wt.$\%$& 40.53 & 620.2 & 1000 & - &536 \ $\pm$ \ 21\\
    Agarose 0.50 wt.$\%$  & 256.34 & 4120.4 & 1000 & -  & 3961 $\pm$ 722\\
    \hline
  \end{tabular*}
  \end{center}
\end{table}

\section{Results and discussion}

In this section we describe liquid jet impacts on liquid and agarose capillary bridges, and compare both cases with previous results of jets impacting liquid pools and droplets. 

In all experiments, upon jet impact on the capillary bridge, a cavity is created. The cavity continues expanding in the x direction with a velocity $U_c$ proportional to the impact speed $U_0$. Furthermore, due the entrained air and momentum conservation this cavity also expands in the radial direction \cite{speirs2018entry}. At a certain instant, the cavity stops expanding, reaches a maximum volume and starts retracting or pinches off. In this process, trapping of air bubbles inside the capillary bridge can occur depending on the impact and liquid characteristics. We then extract the cavity profile and the cavity front position for each frame of the videos. 

In the following sections we will discuss in detail the cavity dynamics. In section III.A, we focus on the expansion of the cavity upon impact and we provide two models and compare with experimental results. Furthermore, we study the critical transition Weber number for the jet to fully traverse or get embedded into the capillary bridge and compare it to jets impacting on pendant droplets.  In section III.B, we discuss the different types of cavity collapse in terms of the liquid and impact characteristics. Finally, in section III.C we study the effect of wettability on the cavity formation and collapse based on the energy dissipation on the contact line.    

\subsection{Model for the cavity expansion}

\begin{figure}
\centering
\includegraphics[width=15cm]{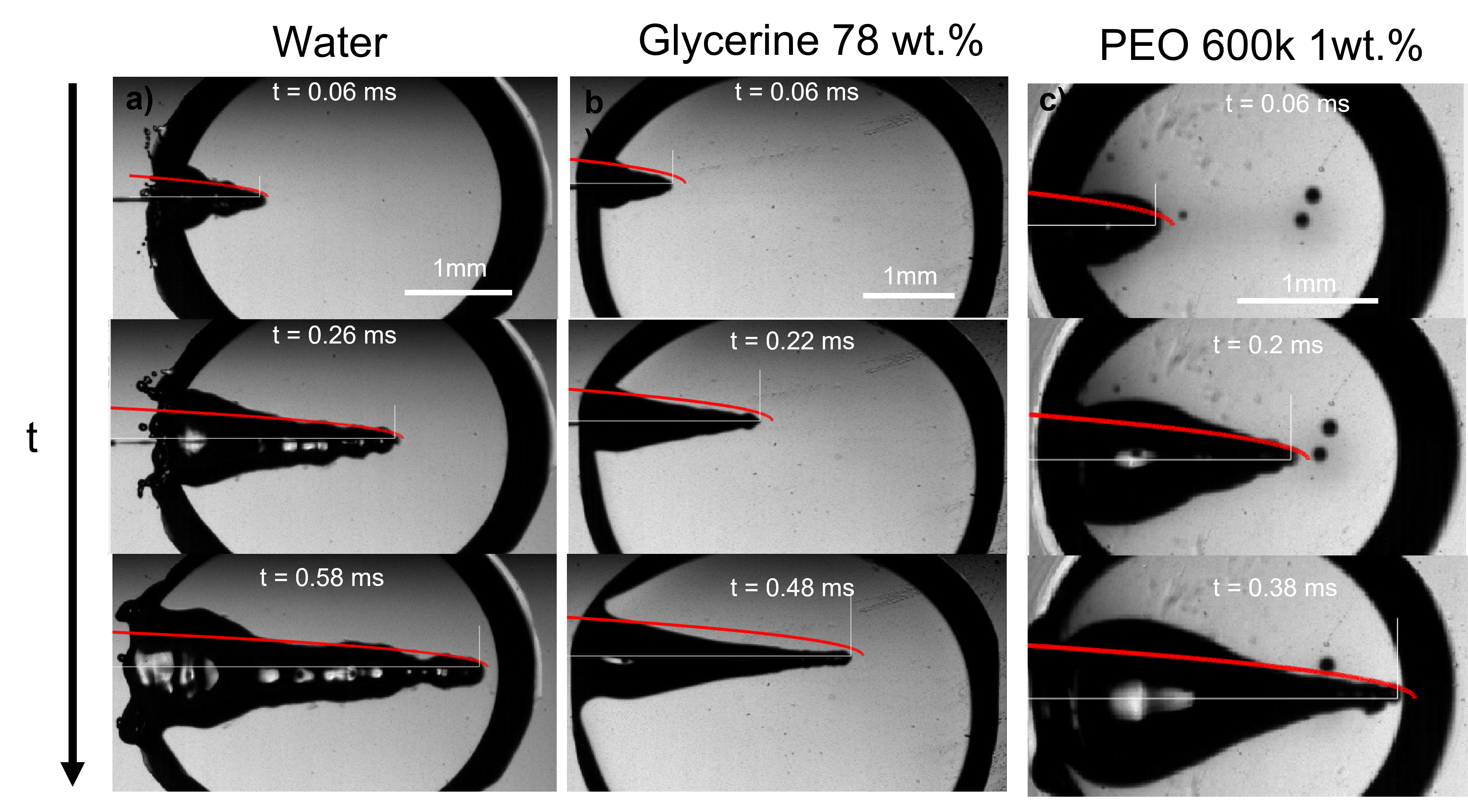}
\caption{Experimental images showing a liquid jet impacting; a) a water capillary bridge, $We_{jet}$ = 495 and $D_{jet} =$ 93 $\mu$m (Movie 1 in the supplementary material), b) a glycerol 78$\%$ capillary bridge  $We_{jet}$ = 483 and $D_{jet} =$ 74 $\mu$m (Movie 2 in the supplementary material) and c) a PEO 600k 1 wt.$\%$ capillary bridge  $We_{jet}$ = 498 and $D_{jet} =$ 80 $\mu$m (Movie 3 in the supplementary material). In all the images a comparison with the theoretical profile given by equation 4, shown as a red line. We adapted the code \cite{lee2021elongated} from for generating this figures. This code evaluates equation 4 at a given time and plots it on top of the experimental images. The videos associated with this figures can be found in the supplementary material.}
\label{Fig2}
\end{figure}

When a capillary bridge is impacted by a jet, a cavity is generated. The cavity keeps expanding until it reaches a maximum depth. If the maximum depth of the cavity is larger than the capillary bridge diameter $D_{cb}$, the jet will traverse the capillary bridge. In our case, the cavity radial expansion is hindered by the walls confinement, depending on the impact velocity. For liquids, a cavity forms even for the smallest jet impact velocity. In contrast, for the agarose gels a critical impact velocity is necessary to generate a cavity, i.e., inertia has to overcome the elasticity of the agarose gel. This critical velocity depends on the Young's modulus of the gel and polymer content \cite{baxter2005jet, mrozek2015relationship}. 

Multiples studies have been dedicated to the cavity expansion and the evolution of its dimensions after a water entry event \cite{bouwhuis2016impact, le2013viscous, deka2018dynamics, speirs2018water, quetzeri2021impact}. At the moment of jet impact, the cavity creation is dominated by inertia ($We >> 1$). By considering the jet inertia in the x direction and that at the beginning of the crater growth $dH_c/dt = U_c \approx U_0/2 $, where $U_c$ is the velocity of the tip of the expanding cavity $H_c$. Therefore $H_c$ is expected to grow linearly with time and the jet inertia is predominantly converted to the momentum of the surrounding liquid \cite{bouwhuis2016impact}. By solving the two dimensional Rayleigh equation in cylindrical coordinates,

\begin{equation}
    \left(R \frac{d^2R}{dt^2} + \left(\frac{dR}{dt}\right)^2\right)\log \frac{R}{R_{\infty}} + \frac{1}{2}\left(\frac{dR}{dt}\right)^2 \approx \frac{ \gamma}{\rho R},
\end{equation}

we can predict the evolution of the cavity radius $R(x,t)$. Since $We >> 1$ and $Re >> 1$, surface tension can be neglected during expansion and $R(t) \approx (t - t_0)^{1/2}$, where $t = H_c/U_c$ and $t_0 = x/H_c$, and the approximate cavity profile is \citep{bouwhuis2016impact},

\begin{equation}
R(x,t) \approx \sqrt{\frac{D_{jet}}{2} (H_c - x)} = \sqrt{\frac{D_{jet}}{2} (U_{c}t - x)}.
\label{CavityProfile}
\end{equation}

Figure \ref{Fig2} shows experimental images of the expanding cavity created after the impact of a microfluidic jet on a capillary bridge of different liquids. The theory considers that the jet has a uniform radius across it length, therefore a perfectly parabolic profile is expected. However, as the jets in our experiments have a bigger head than the following liquid, and it breaks up before the collapse of the cavity, the cavity radius at the impact point ($R(0,t)$) is underpredicted for all cases. 

Figure \ref{Fig2}a, shows the comparison between the theoretical profile and experimental profile after the impact on a water capillary bridge. The theoretical profile fits the shape of the cavity generated in water, albeit, the cavity is not as smooth as predicted by the theory. This difference arises from assuming a perfectly cylindrical and steady jet in the theory, which is not the case for the experiment. In reality the jet breaks up and subsequent droplets impact the base of the cavity generating surface waves across all the cavity \cite{speirs2018entry} (see Movie 1 in the supplementary materials). 

For the glycerol 78 wt.$\%$ capillary bridge we observe that R(x,t) is overestimated (figure \ref{Fig2}b), as the model does not take into account the viscosity of the capillary bridge and a purely inertial event is considered. To include the viscous dissipation in the cavity formation we start the derivation from the cavity expansion formed after a single droplet impact on a pool. By using potential flow theory and considering shear stresses in a thin layer covering the interface of the cavity, the cavity evolution can be expressed by the following system of equations \cite{bisighini2010crater}, 

\begin{align}
    \frac{d^2R}{dt^2} & = -\frac{3(\frac{dR}{dt})^2}{2r} - \frac{2}{r^2 We} + \frac{7(\frac{dx}{dt})^2}{4r} - \frac{(4\frac{dR}{dt})}{r^2 Re} 
    \\ 
    \frac{d^2x}{dt^2} & = -\frac{3(\frac{dR}{dt})(\frac{dx}{dt})}{r} - \frac{9(\frac{dx}{dt})^2}{2r} - \frac{12\frac{dx}{dt}}{r^2 Re}.
\end{align}

Similarly to Speirs et al. 2018 \cite{speirs2018water}, we use the resemblance of a cavity generated by a train of droplets and liquid jet. We use the solution of the cavity expansion of a single droplet given by equations 5 and 6, and at a given time we superimpose the solution of the expansion of the next impacting droplet. This allows us to predict the cavity evolution produced by a jet in a viscous capillary bridge. To generate a comparable cavity to that of a jet, the train of droplets needs to have a similar volumetric flux as a jet. Therefore, we equate the volume of a droplet $V_d = \frac{4}{3}$ $\pi r_d^3$ to that of a cylinder that encapsulates a train of droplets, $V_{cyl} = \pi r_{cyl}^2 h_{cyl}$, 

\begin{equation}
    r_d = r_{cyl} \sqrt{\frac{3}{2f}},
\end{equation}

where $h_{cyl} = 2r_d/f$ is the droplets centre-to-centre distance, $f$ is the frequency of the train of droplets and $r_{cyl}$ is the cylinder radius.  
Furthermore, previous work show that the cavity velocity is dependent to the frequency of the droplet train \cite{bouwhuis2016impact},

\begin{equation}
    U_c = \frac{\sqrt{f}}{1 + \sqrt{f}} U_d,
\end{equation}

where $U_d$ is the droplet velocity. Therefore, for a train of droplets to have a similar cavity velocity than for a jet with velocity $U_0$ we need,

\begin{equation}
    U_d = \frac{2\sqrt{f}}{1 + \sqrt{f}} U_0.
\end{equation}

\begin{figure}
\centering
\includegraphics[width=15cm]{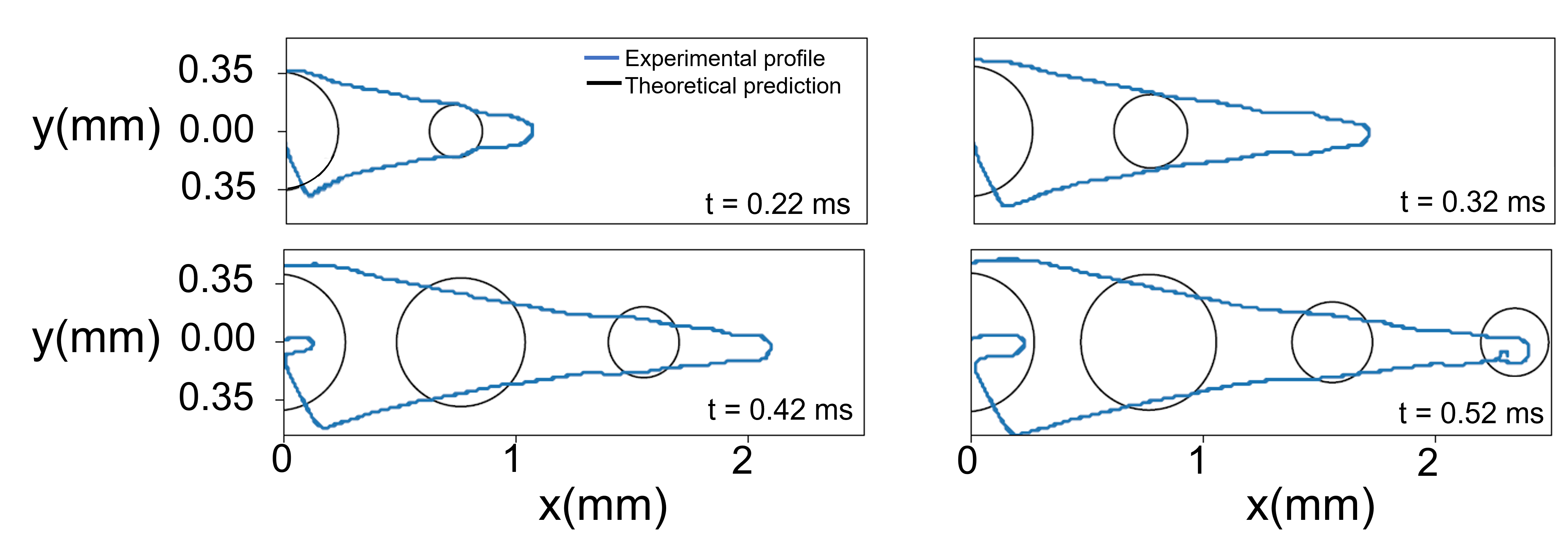}
\caption{Comparison between the experimental cavity profile (blue line) and the theoretical profile for the droplet train (black circles). The experimental conditions are the same as for the impact on Glycerine 78 $\%$ in figure \ref{Fig2} }
\label{Fig3}
\end{figure}

We take the corrections of equations 7 and 9 for a train of droplets with a frequency f = 1/2, and overlaying it to the experimental profiles of glycerol 78$\%$ and show the results in figure \ref{Fig3}. As observed in figure \ref{Fig3} the agreement of the expansion at times t = 0.32ms and t = 0.72ms is excellent. In contrast, the expansion radius r(x) is slightly over-predicted for times t =1.12ms and t = 1.52 ms. However, overall the model of the train of droplets provides a better estimate of r(x) than the use of the two-dimensional Rayleigh equation, where the viscous dissipation is neglected. 

For the PEO 600k 1wt.$\%$ the theoretical cavity profile fits well with the prediction in equation \ref{CavityProfile} close to the tip of the cavity (figure \ref{Fig2}). The cavity also shows, a smoother profile compared to that of water. The latter can be expected as PEO 600k 1wt.$\%$ has a viscosity $\approx$ 20 times higher than water, damping the capillary waves.

In addition to studying the cavity geometry during expansion, we tested the model for the critical velocity of a jet traversing a liquid droplet that we developed previously \citep{quetzeri2021impact}. Assuming that the surface tension dominates during the cavity collapse and using the shape of the cavity described by equation \ref{CavityProfile} we can predict the time of the collapse of the cavity \cite{quetzeri2021impact},

\begin{equation}
    t_c \approx \frac{\rho_0^2 D_{0}^3 U_c^3}{512 \gamma_{cb}^2}.
    \label{t_c}
\end{equation}

Then if the diameter of the capillary bridge $D_{cb}$ is larger than $H_{max} = (U_0/2) t_c$ the jet would traverse the droplet. Rearranging so that we have a critical Weber number for the traversing, we obtain,

\begin{equation}
    We^* \approx \left(\frac{2^{13} D_{cb}}{D_{0}}\right)^{1/2}.
    \label{WeCriticalLaplace}
\end{equation}

To compare this result with the experimental data, we use the ratio between the critical traversing Weber number $We^*$ and the experimentally measured one $We$, i.e.,

\begin{equation}
    \frac{We}{We^*} = \frac{\rho_{0} U_{0}^2 D_{0}^{3/2}}{(2^{13/2}) \gamma_{cb} D_{cb}^{1/2}}.
    \label{GammaDyn}
\end{equation}

\begin{figure}
\centering
\includegraphics[width=8cm]{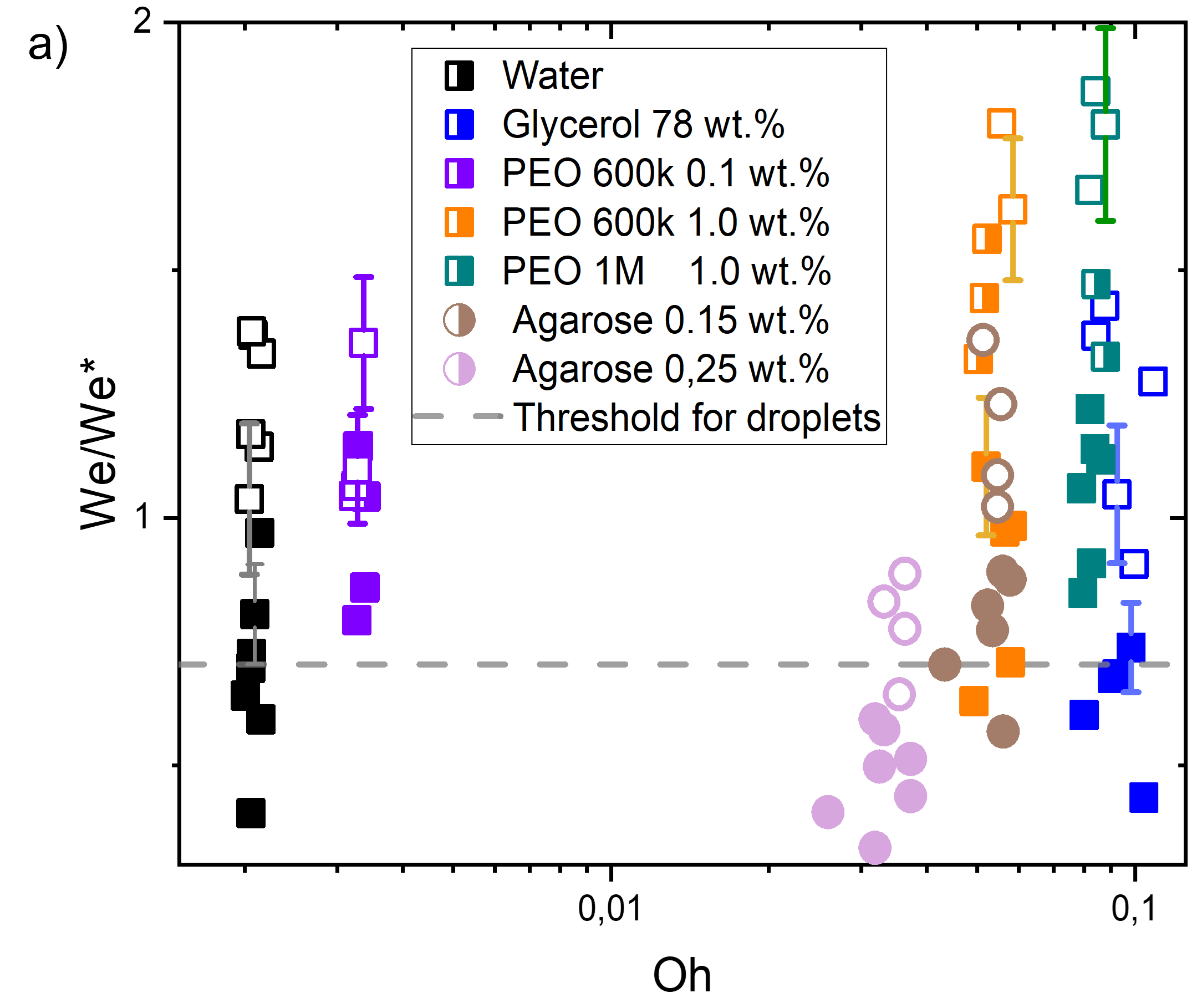}
\includegraphics[width=8cm]{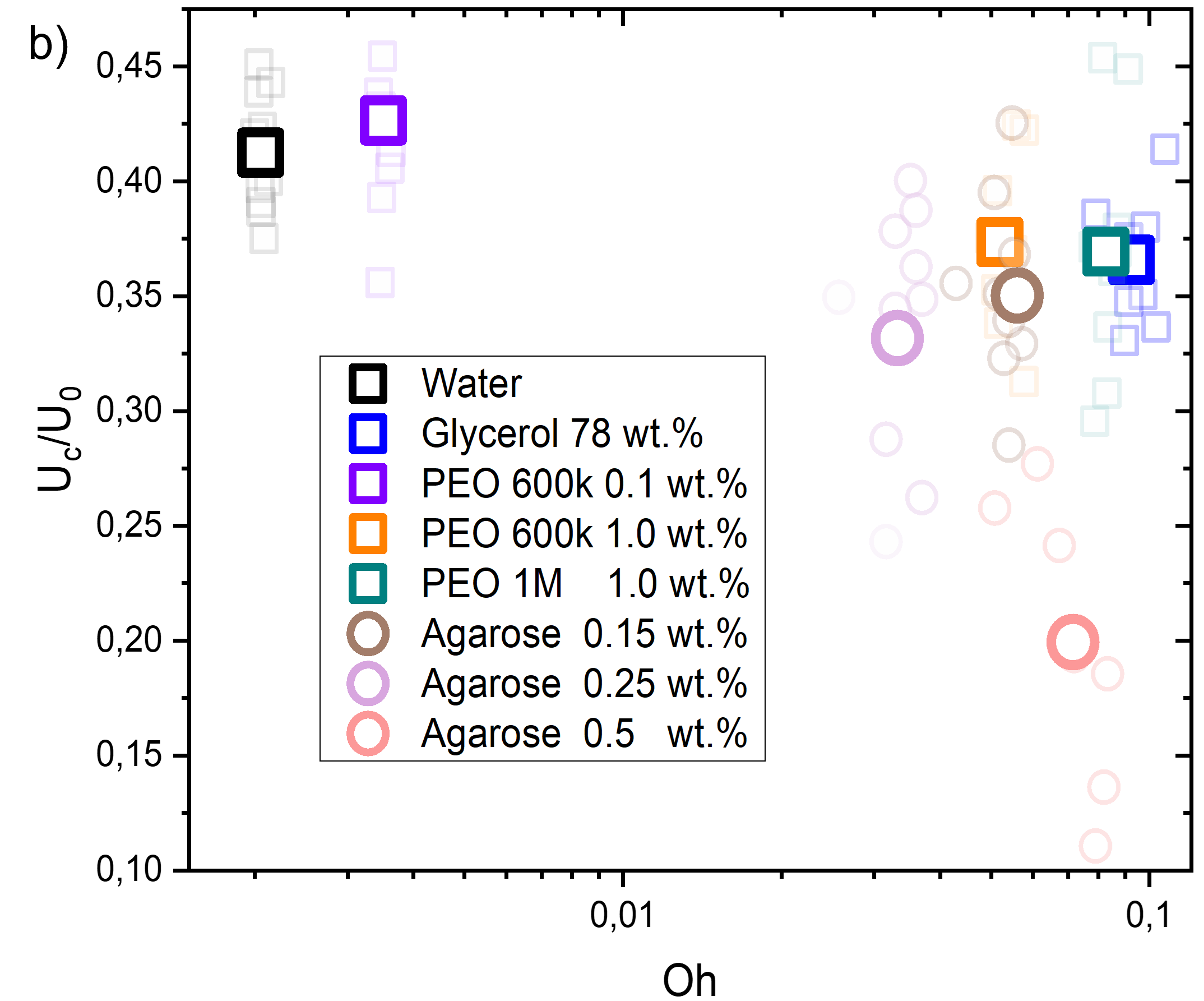}
\caption{a) Phase diagram comparing the critical Weber number for traversing obtained experimentally (We) and from equation 11 ($We^*$). Each liquid and agarose gel are represented by their respective Oh number. Open symbols show cases where the jet traverses the droplet, while solid ones stand for the embedding case. The dashed line is $We/We^* = 0.7$ the threshold found for droplets in \cite{quetzeri2021impact}. The model is in good agreement with the experimental data for liquids and agarose 0.15 $\%$ wt. In contrast, the threshold is underestimated for the agarose 0.25 $\%$ wt and 0.5 $\%$ wt (not shown due to scale). Uncertainty was calculated for all the experimental data and example error bars are shown at selected points. b) Ratio of the cavity velocity $U_c$ and the impact velocity $U_0$, for the liquids and agarose gels studied in this experiments. Even for liquids with Oh $<$ 0.01, $U_c/U_0$ is lower than the predicted 0.5, we attribute this to the wall dissipative effects.}
\label{Fig4}
\end{figure}

Figure \ref{Fig4}a shows the comparison between the threshold predicted by equation \ref{GammaDyn} and the threshold found in our experiments. The Ohnesorge number ($Oh = \mu_{cb} /\sqrt{\rho_{cb} D_{cb} \gamma_{cb}}$, with subscript cb indicating the capillary bridge properties) represents the ratio between viscous to inertial and capillary forces. The threshold for traversing the Newtonian capillary bridges is within 10 $\%$ of the model prediction. In contrast, for the PEO 600k and 1M 1.0 wt $\%$ the threshold deviates in $\approx$ 50 $\%$, as the liquid viscoelasticity influences the process \cite{quetzeri2021impact}. 

In figure \ref{Fig4} the dashed line shows the experimental threshold between embedding and traversing for the impact of a microfluidic jet on pendant droplets \cite{quetzeri2021impact}. We find that the threshold for traversing Newtonian capillary bridges is higher than for pendant droplets. This is consequence of assuming that the cavity advances at a constant velocity $U_c = 1/2\,U_0$. However, as shown in figure \ref{Fig4}b, $U_0 < 1/2\,U_c$ for all the capillary bridges. We attribute this decrease in $U_c$ to the confinement, provided by the glass holding the capillary bridge. Indeed, in previous experiments the influence of confinement has been found to reduce the cavity radial expansion and increase the energy dissipation \cite{zou2013phenomena, bouchard2019droplet, guo2020investigation,primkulov2020characterizing}. 
Moreover, in figure \ref{Fig4}b we show that the velocity of the expanding cavity depends on both the viscosity and density ratios of the target and projectile. By considering viscous dissipation ($\mu U_0^2 / D_0^2$), and balancing the initial kinetic energy of the jet with the displaced liquid on the pool and energy of the jet at a time t after impact. Then, $U_c$ can be expressed as \cite{fudge2021dipping}, 

\begin{equation}
    U_c \approx \frac{1}{\sqrt{1 + 2.71 \rho_r + (24/Re) \mu_r}}, 
\end{equation}

where $\rho_r$ and $\mu_r$ are the density and the viscosity ratios between the microfluidic jet and the capillary bridge. 


For agarose 0.15 wt.$\%$, the prediction of the traversing threshold is within the same range of values than for the liquids. For agarose  0.25 wt.$\%$, the threshold is of the same order of magnitude than for liquids, but differs by $\approx$ 30 $\%$. Furthermore, the threshold found for agarose 0.5 wt.$\%$ is one order of magnitude lower than the predicted value (not shown in the plot). The average measured $U_c/U_0$ for agarose 0.5 wt.$\%$, is five times lower than the one used to predict the traversing, however this difference is insufficient to explain the discrepancy.

That the predicted transition occurs at a lower $We/We^*$ for increasing agarose concentration seems contradictory as for increasing elastic modulus there is an increase in viscous dissipation \cite{shao2020method}. This underestimation of the traversing threshold could be attributed to an overestimation of the surface tension for agarose gels with G $\gtrsim$ 500 Pa. The linear relations used for calculating the surface tension and viscosity were obtained for agarose gels with concentrations lower than 0.25 wt.$\%$ and with G $<$ 200 Pa \cite{shao2020method}. Therefore we see a good agreement with the agarose 0.15 wt.$\%$ (which G $=$ 176) , but possibly for gels with G $>$ 500 Pa the linear relation is no longer valid. We expect that this results translate to skin or other complex materials, where surface tension and viscosity are difficult to asses and define. 

\subsection{Cavity collapse}

\begin{figure}
\centering
\includegraphics[width=12cm]{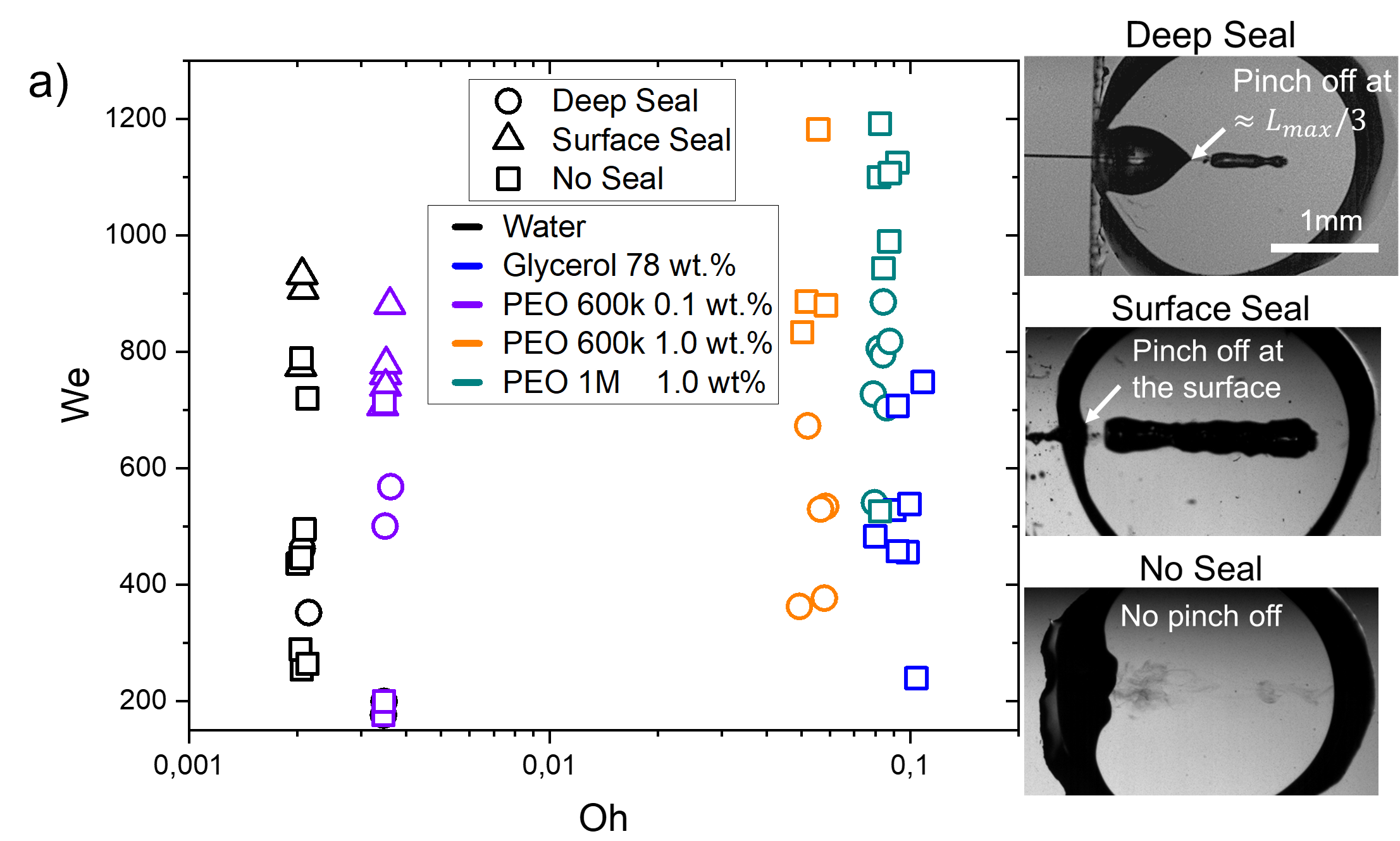}
\includegraphics[width=8cm]{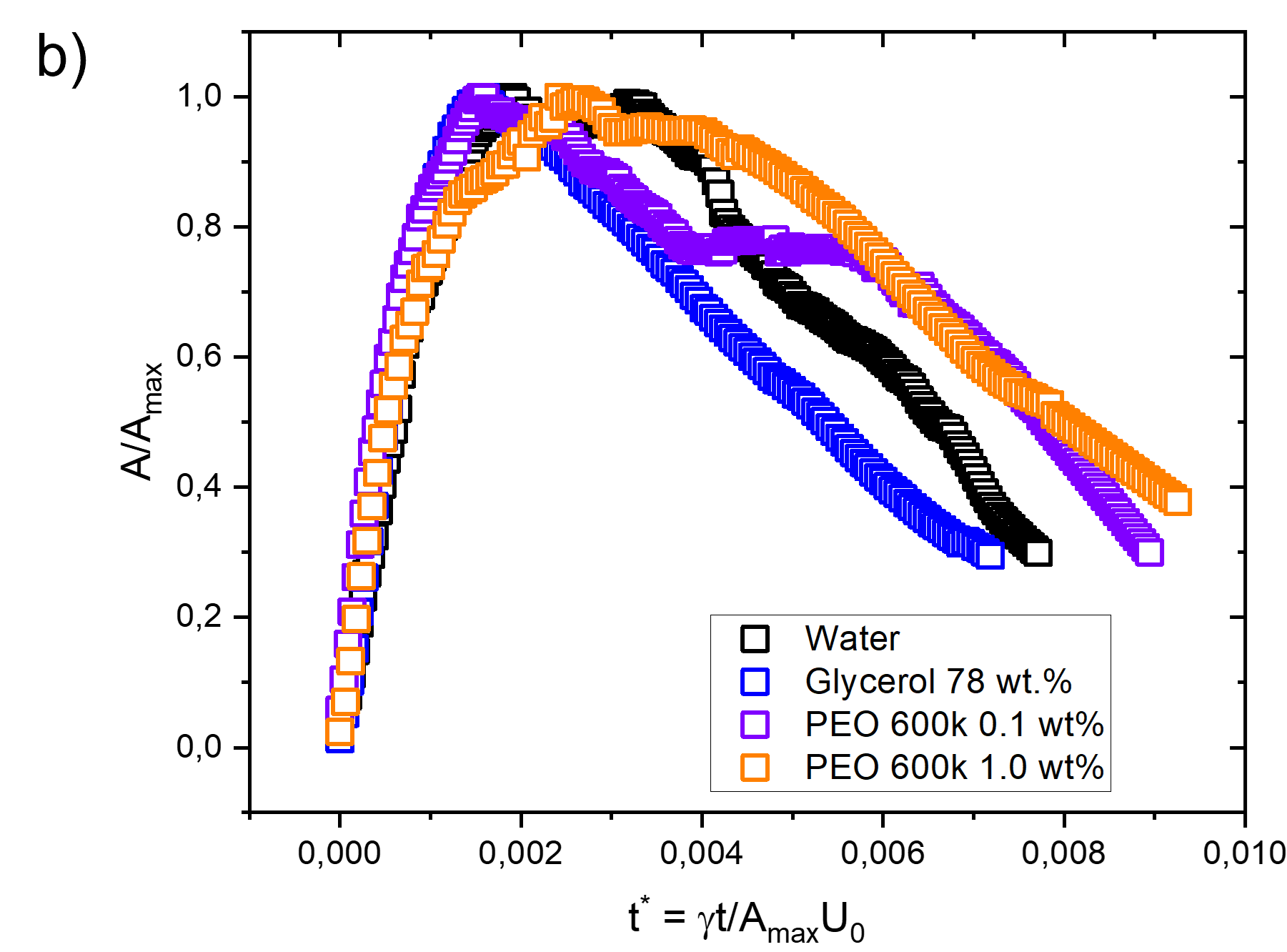}
\caption{a) Different types of seals as a function of $We_{jet}$ and $Oh$. The material properties influence the type of collapse observed at the same Weber numbers. Experimental snapshots of the different type of closure are shown on the right. Movies 1,2 and 3 in the  supplementary material show the `no seal' regime. Movies 4 and 5 represent `deep seal' and `surface seal' regimes respectively . b)Normalised area of the cavity $A/A_{max}$ of different liquids in terms of non-dimensional time $t^* = \gamma_{cb} t/ A_{max}U_0$. The cavity collapse is dependent on the cavity maximum area, the impact velocity and the surface tension of the liquid. Furthermore, the viscoelastic response of the PEO solutions resist the cavity collapse, retarding the collapse as compared to glycerol 78 $\%$ and water. The impact conditions are the same than the cases in figure \ref{Fig2}. }
\label{Fig5}
\end{figure}

\begin{figure}[hbt!]
\centering
\includegraphics[width=8cm]{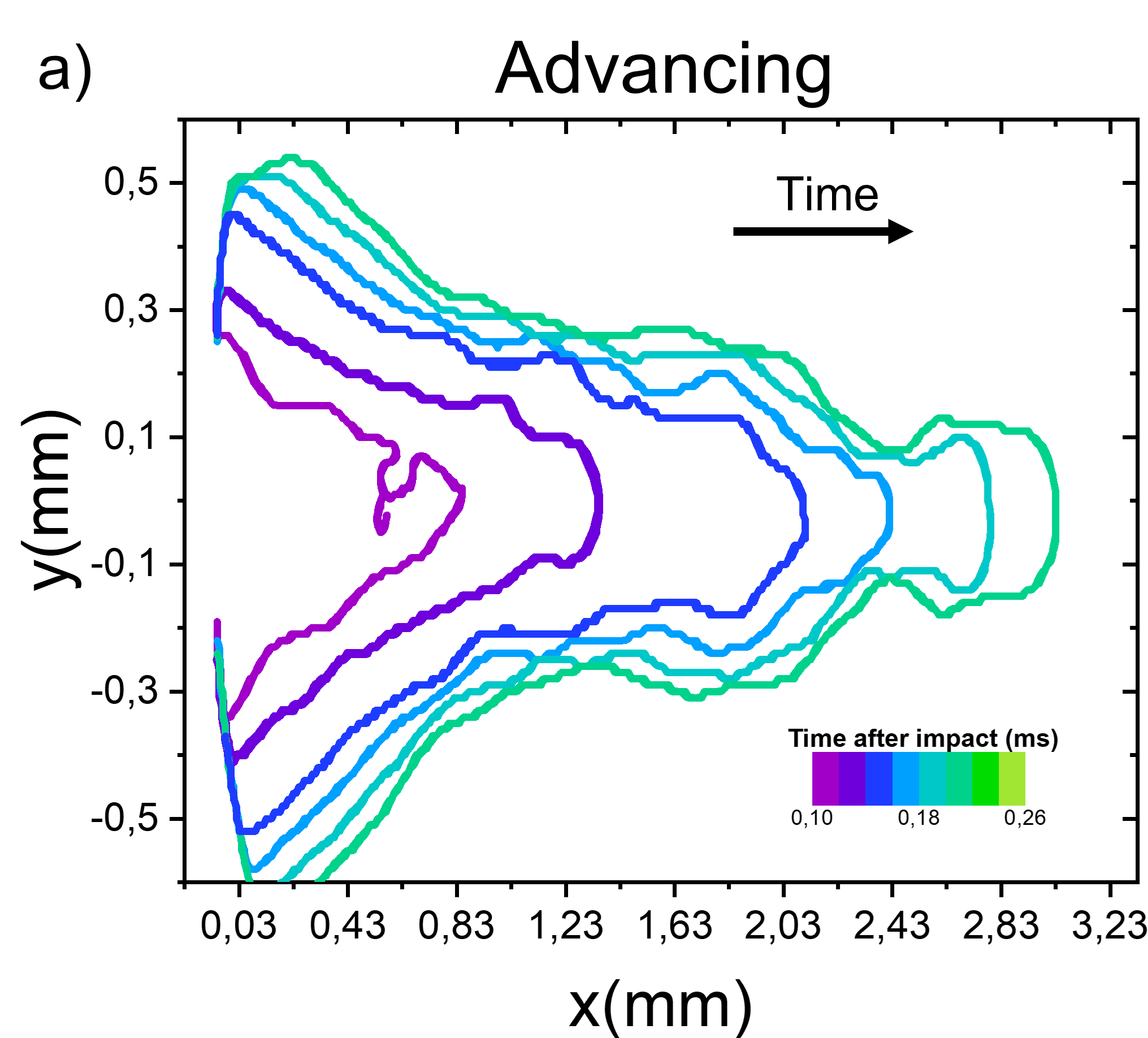}
\includegraphics[width=8cm]{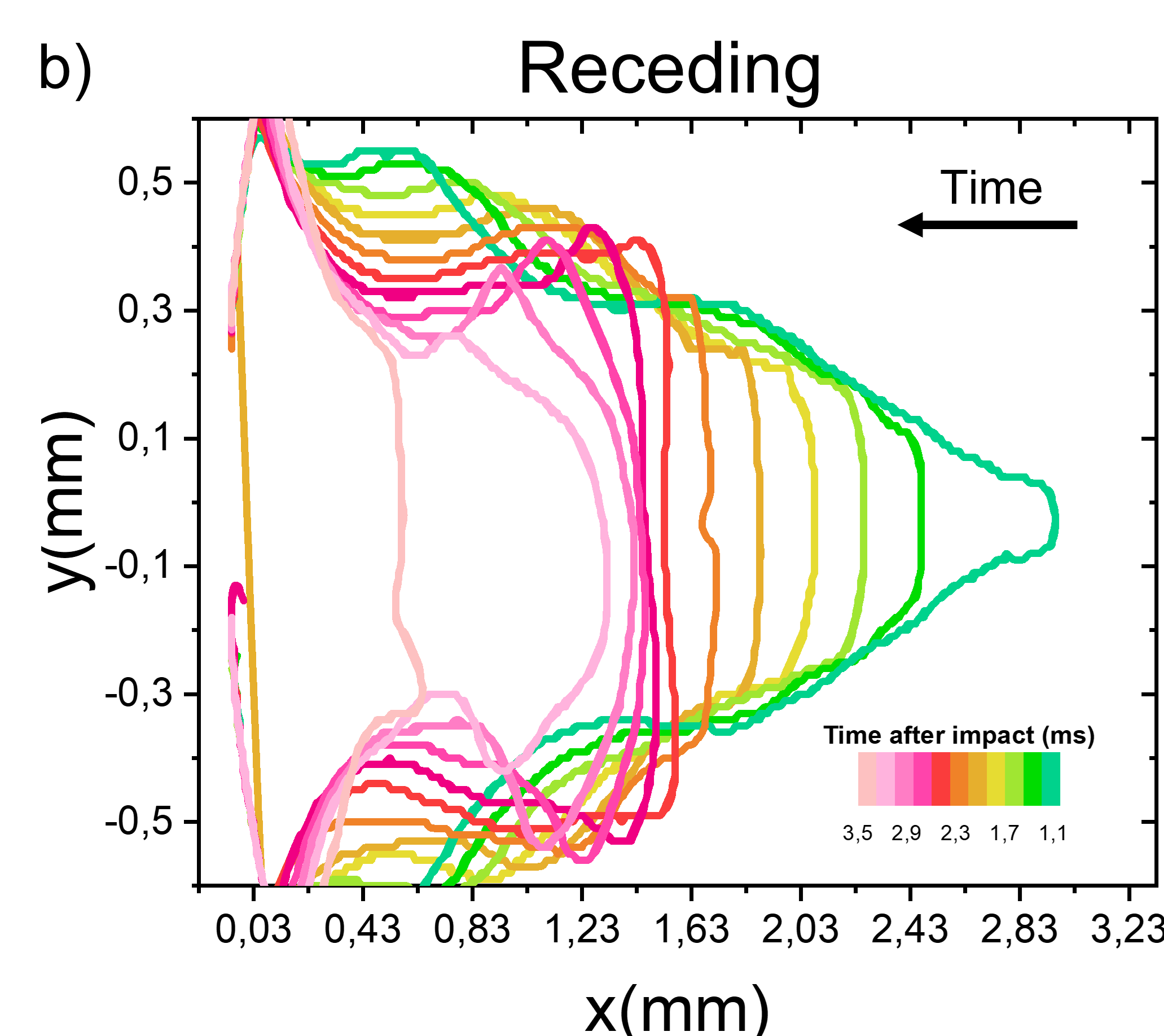}
\includegraphics[width=8cm]{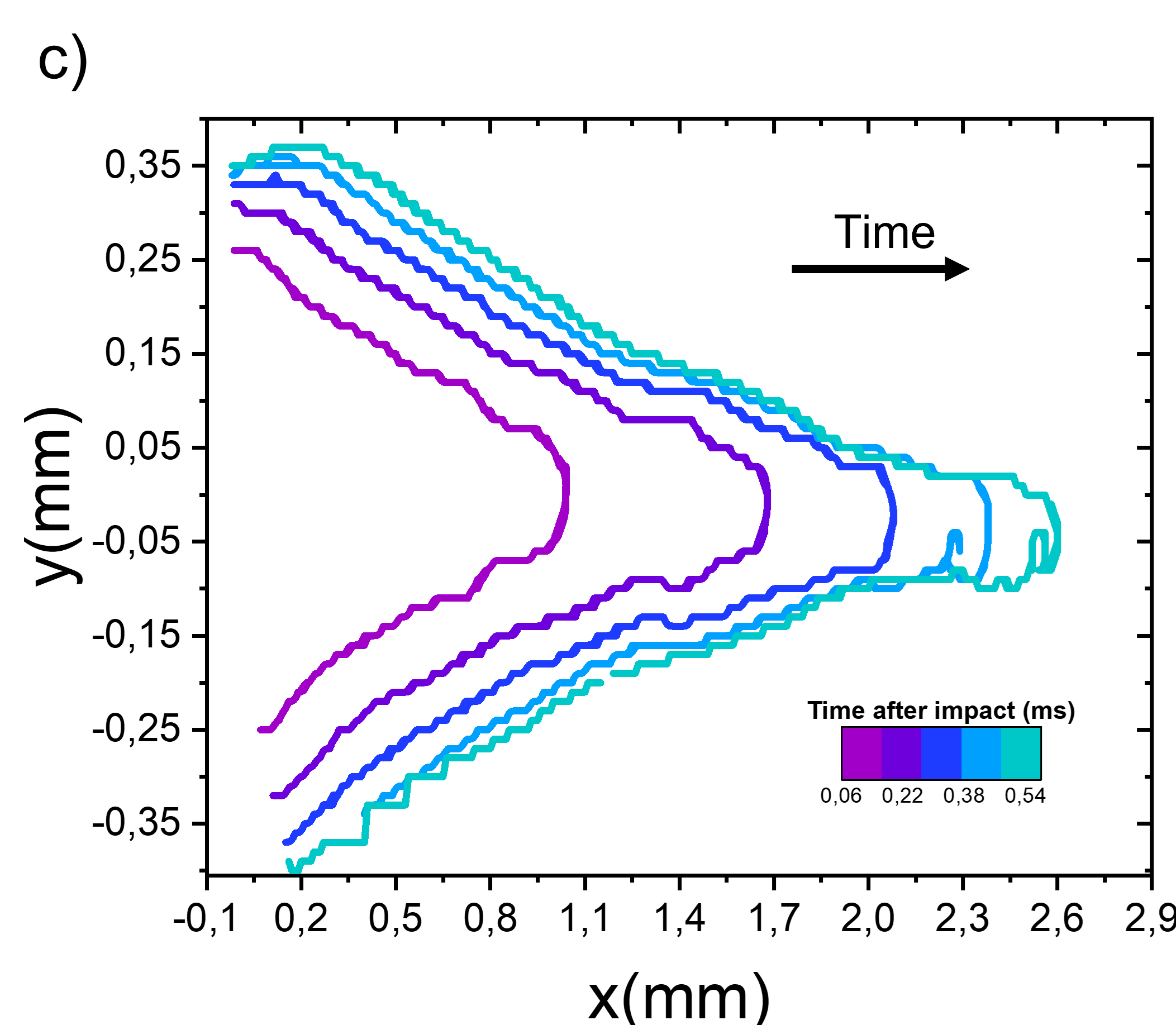}
\includegraphics[width=8cm]{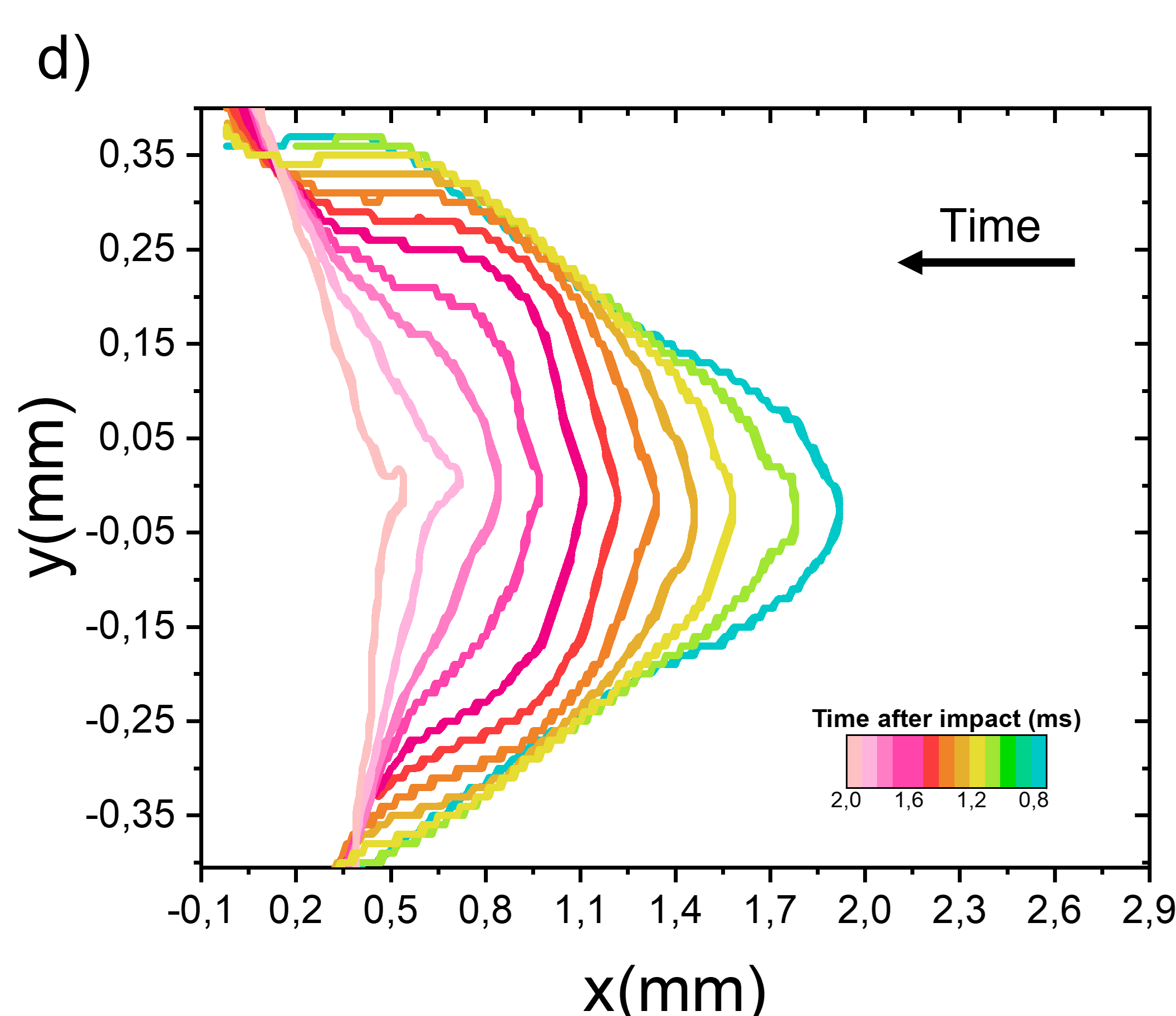}
\includegraphics[width=8cm]{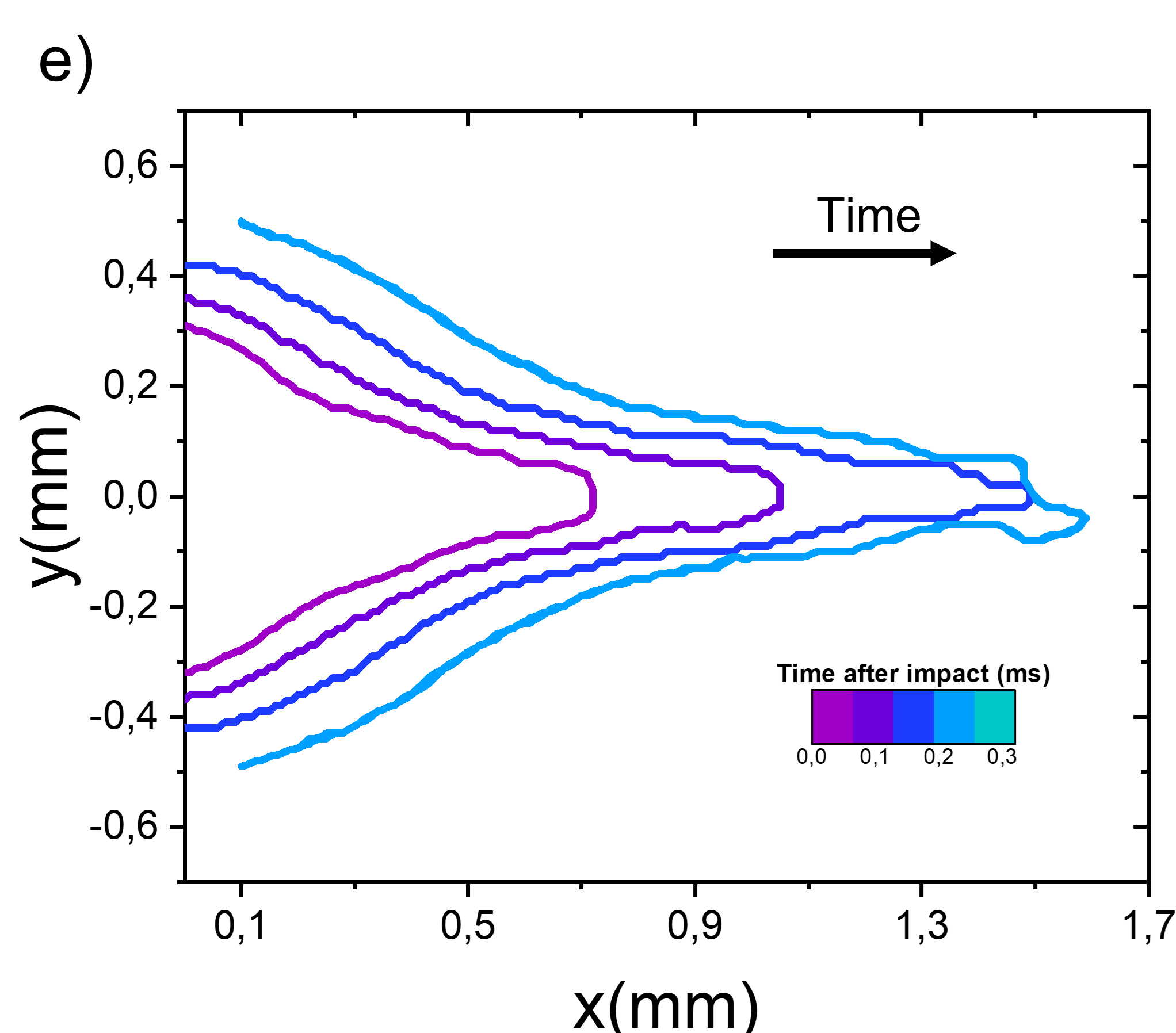}
\includegraphics[width=8cm]{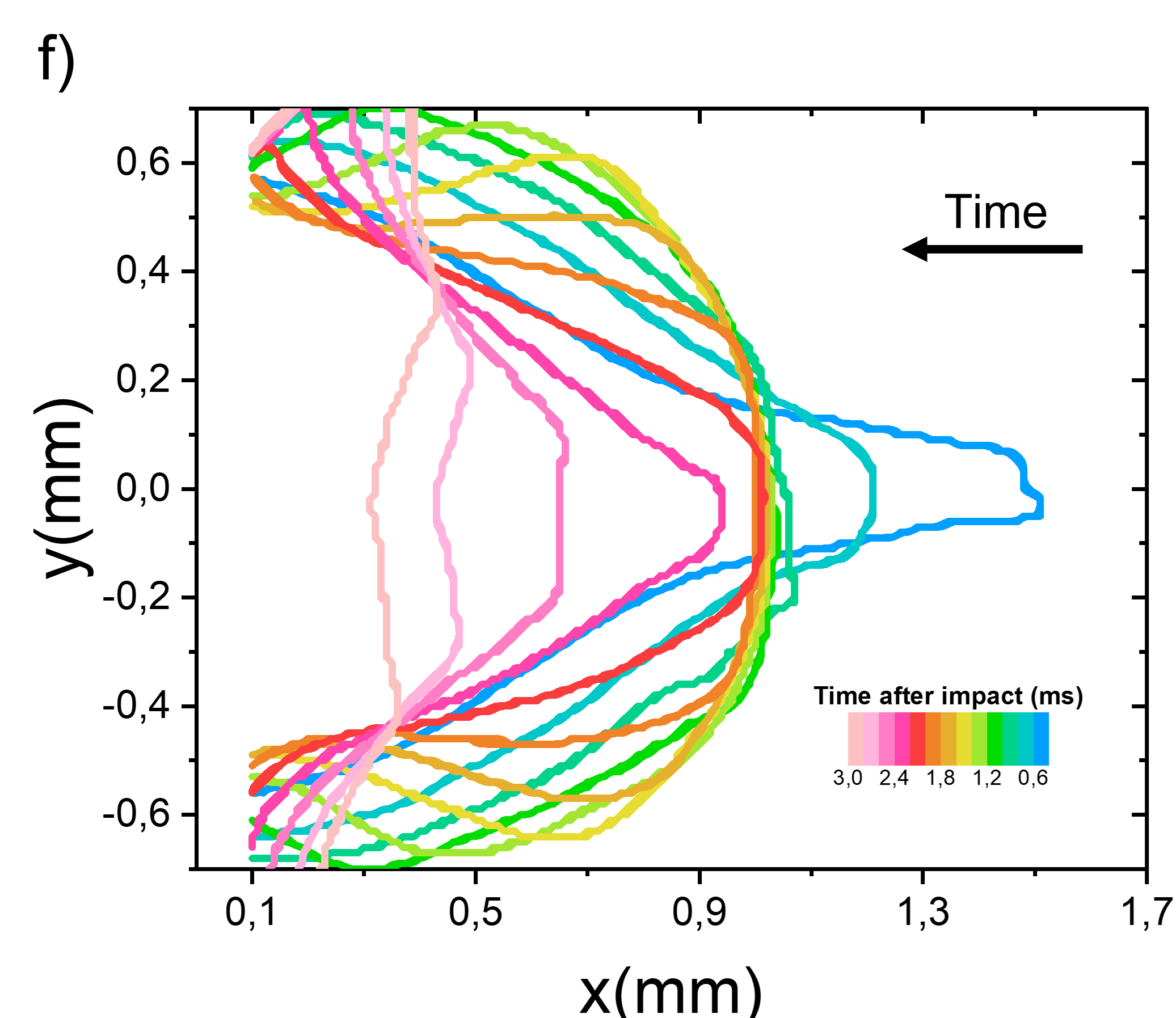}
\caption{Cavity profiles after the impact of a microfluidic chip on a capillary bridge. a),b) Water capillary bridge; we observe capillary waves travelling across the cavity throughout the process. c),d) Glycerol 78 wt. $\%$. capillary bridge; the capillary waves are damped due to viscous dissipation. e),f) PEO 600k 1 wt. $\%$ capillary bridge; while the cavity is receding the cavity front stops from $ t\approx$ 1.8 ms to $t \approx$ 2.4 ms the tip of the cavity stays roughly at the same position. The experimental conditions are the same as in figure \ref{Fig2}}
\label{Fig6}
\end{figure}

\subsubsection*{Closure regimes}

We classify the closure type of the cavities in four categories: `no seal', `deep seal', `shallow seal' and `surface seal' as shown in figure \ref{Fig5}b \cite{aristoff2009water,kiyama2019gelatine}.  The type of seal depends on the capillary bridge properties as well as on the Weber number.

In the `no seal' regime, the cavity stops at $L_{max}$ or at $D_{cb}$ in case of traversing and retracts until the capillary bridge returns to a circular shape. In addition, we define the `deep seal' regime as the phenomena where the cavity collapses at a pinch point larger than $L_{max}/3$. Furthermore, within the `shallow seal' regime, the cavity collapses close to the jet entry point. Moreover, the `surface seal' regime, refers to the closure of the surface when the cavity is still expanding, and a overarching dome that closes at a location $x<x_0$, where $x_0$ is the impact point. 

During the `no seal' regime, in contrast to the impact of droplets on liquid pools or a jet on a pendant droplet, no Worthington jets are observed. We attribute this to the reflection of the surface waves on the glass walls as it was found that capillary waves control the cavity pinch-off and Worthington jet properties \cite{michon2017jet,deka2018dynamics, deike2018dynamics}. Furthermore, at the base the cavity can expand until contacting the walls. This causes the liquid to wet an area larger to that of the initial capillary bridge. Glass is wettable and therefore the contact line gets pinned into the walls that contain the liquid capillary bridge. Thus, the cavity retraction suffers from energy loses due to viscous and contact line dissipation and a Worthington jet is no longer energetically favourable. A more detailed discussion on the wettability effects of the walls is presented in Section III.C. Furthermore, for this regime to occur the aspect ratio of the cavity at the onset of retraction, the length of the jet and the We number are crucial parameters to consider \cite{deka2018dynamics}. In this type of retraction no bubbles are trapped inside the capillary bridge.
In the `deep seal' regime, after the cavity pinches-off, the cavity may further pinch-off at multiple locations. Therefore, multiple bubbles and antibubbles are trapped inside the capillary bridge. Where antibbubles are defined as droplets with an air shell \cite{vitry2019controlling}. We observed this type of sealing regime for an increasing Weber number with an increasing PEO molecular weight and concentration. For PEO 600k 0.1 wt. $\%$ deep seal is observed for $We \approx 500-600$, for PEO 600k 1.0 wt.$\%$ $We \approx 400 - 700$ and for PEO 1M 1.0 wt.$\%$ $We \approx 500 - 900$. For increasing PEO molecular weight and concentration the PEO solutions are more viscous and have a larger relaxation time. Consequently, cavities are smaller at the same $We$ for increasing molecular weight and concentration. Furthermore, we observe that viscoelasticity is crucial for the `deep seal' regime to occur. Indeed, glycerine 78 $\%$ has a similar $Oh$ as compared to PEO 1M 1.0 wt. $\%$, but is a Newtonian liquid and this type of seal is not observed.
Within the `shallow seal' regime, the cavity collapses close to the jet entry point. This type of seal was observed for all the agarose capillary bridges at all the explored conditions, while it was absent in Newtonian and viscoelastic capillary bridges. This regime is characterised by the entrapment of several bubbles.
The `surface seal' regime, refers to the closure of the surface when the cavity is still expanding, and a overarching dome that closes at a location $x<0$. The latter is observed for $We \gtrsim 700$ and low viscosity liquids ($Oh \lesssim 0.004$). This regime has been studied in detail in refs. \cite{aristoff2009water, eshraghi2020seal}.

\subsubsection*{Cavity profiles}

Here, we study how the area of the cavity during the collapse stage is affected by the liquid properties and the cavity profiles at different times. In figure \ref{Fig6}  we show the cavity profiles of different liquids during advancing and receding for We $\approx$ 490. For the water capillary bridge, propagating surface waves can be observed for both the advancing and receding of the cavity (figures \ref{Fig6} a and b). In contrast, for the aqueous glycerol mixture capillary bridge, the propagating waves are damped by the viscosity of the liquid \cite{jenkins1997wave}. Furthermore, for the PEO600k wt. 1$\%$ capillary bridge we observe that the front of the cavity stops at an instant comparable to the polymer relaxation time ($t \approx \lambda \approx 1.5$ ms). This retardation in the collapse has been observed for voids generated in a viscoelastic fluid and it was attributed to the liquid elasticity \cite{fogler1970collapse}. 

Figure \ref{Fig6}, also shows that the $R(x,t)$ is the smallest for the glycerine $78\%$ this can be explained from equations as discussed in Section III A. Furthermore, the PEO 600k 1 wt.$\%$ has the smallest $L_{max}$. This is inline with previous work where it was found that viscoelastic liquid, make a bullet decelerate faster than purely viscous or shear-thickening liquids with the same zero shear viscosity  \cite{de2019high}. 

Additionally, we show in figure \ref{Fig5} the normalised area of the cavity $A^*(t) = A/A_{max}$, where $A_{max}$ is the maximum area of the cavity, in terms of the non-dimensional time $t^* = \frac{\gamma_{cb} t}{\rho_{c} U_{jet}A_{max}^2}$, for the same impact conditions as in figure \ref{Fig6}. This non-dimensional time makes all the cavities comparable, as even the same We number can lead to a different $A_{max}$ for different liquids and the time for collapse is strongly dependent on this parameter \cite{deka2018dynamics}. For example, viscosity makes cavity expansion less efficient as discussed in section III A. Therefore the cavities created in the Glycerol 78 wt.$\%$ are smaller than for water at similar We numbers. As observed in figure \ref{Fig5}, the cavity collapse happens at the same $t^*$ (within the experimental error) for water an glycerol 78 wt.$\%$. In contrast, due to their viscoelastic behaviour \cite{fogler1970collapse}, the PEO solutions the cavity collapse is delayed by up to $\approx 25 \%$  as compared to water.

\subsubsection*{Entrained bubbles}

\begin{figure}
\centering
\includegraphics[width=4.6cm]{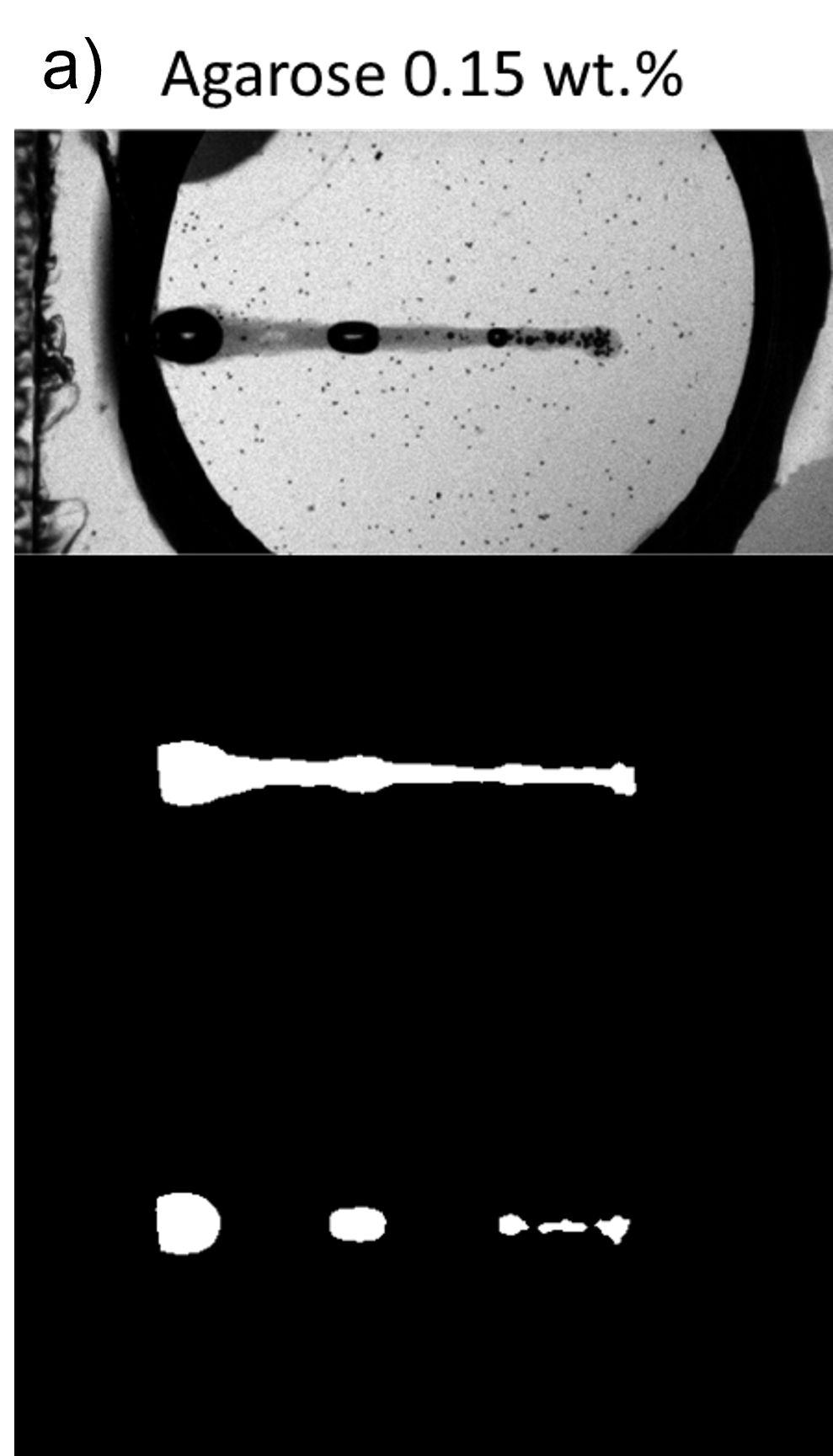}
\includegraphics[width=6cm]{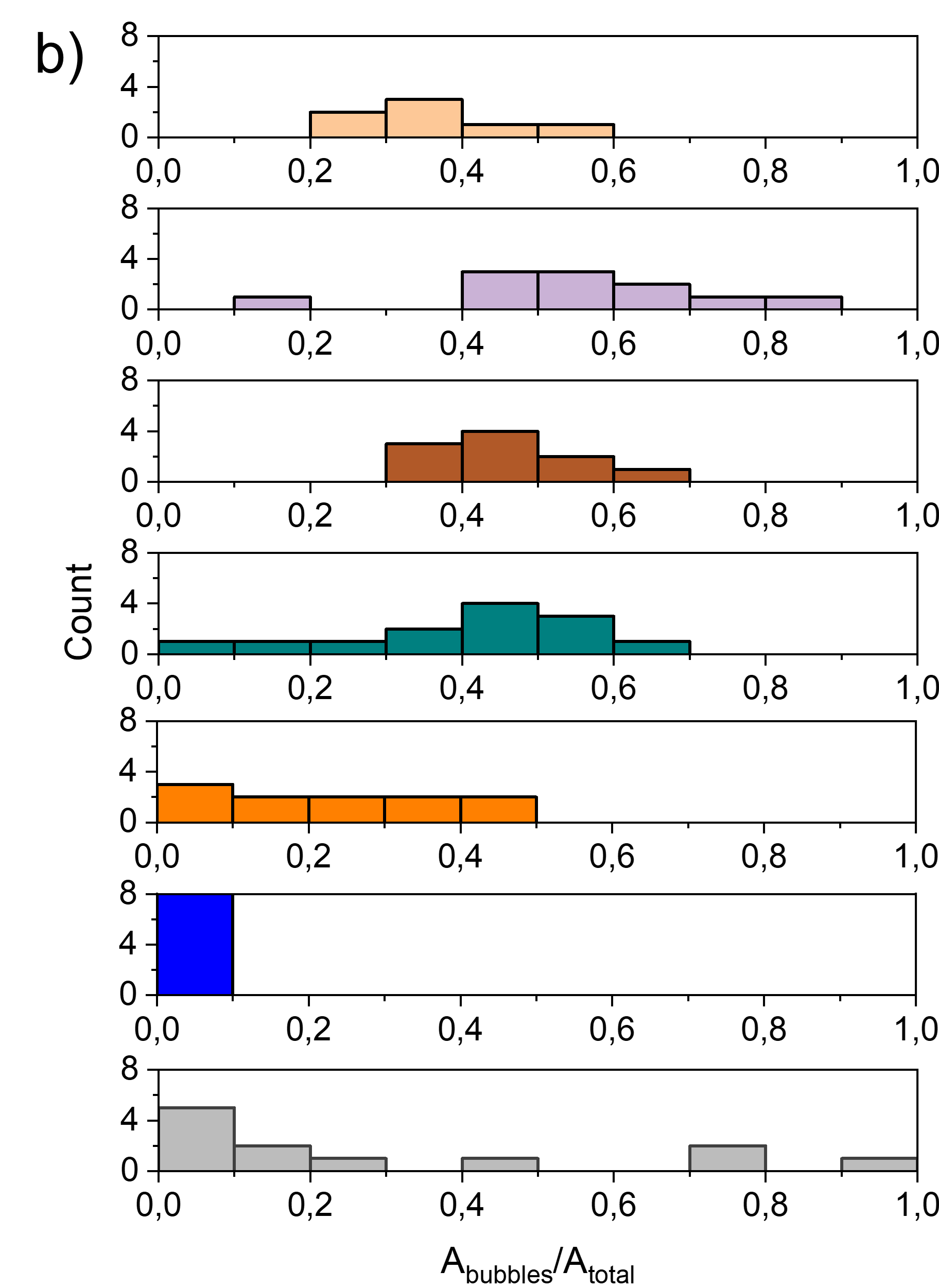}
\includegraphics[width=6cm]{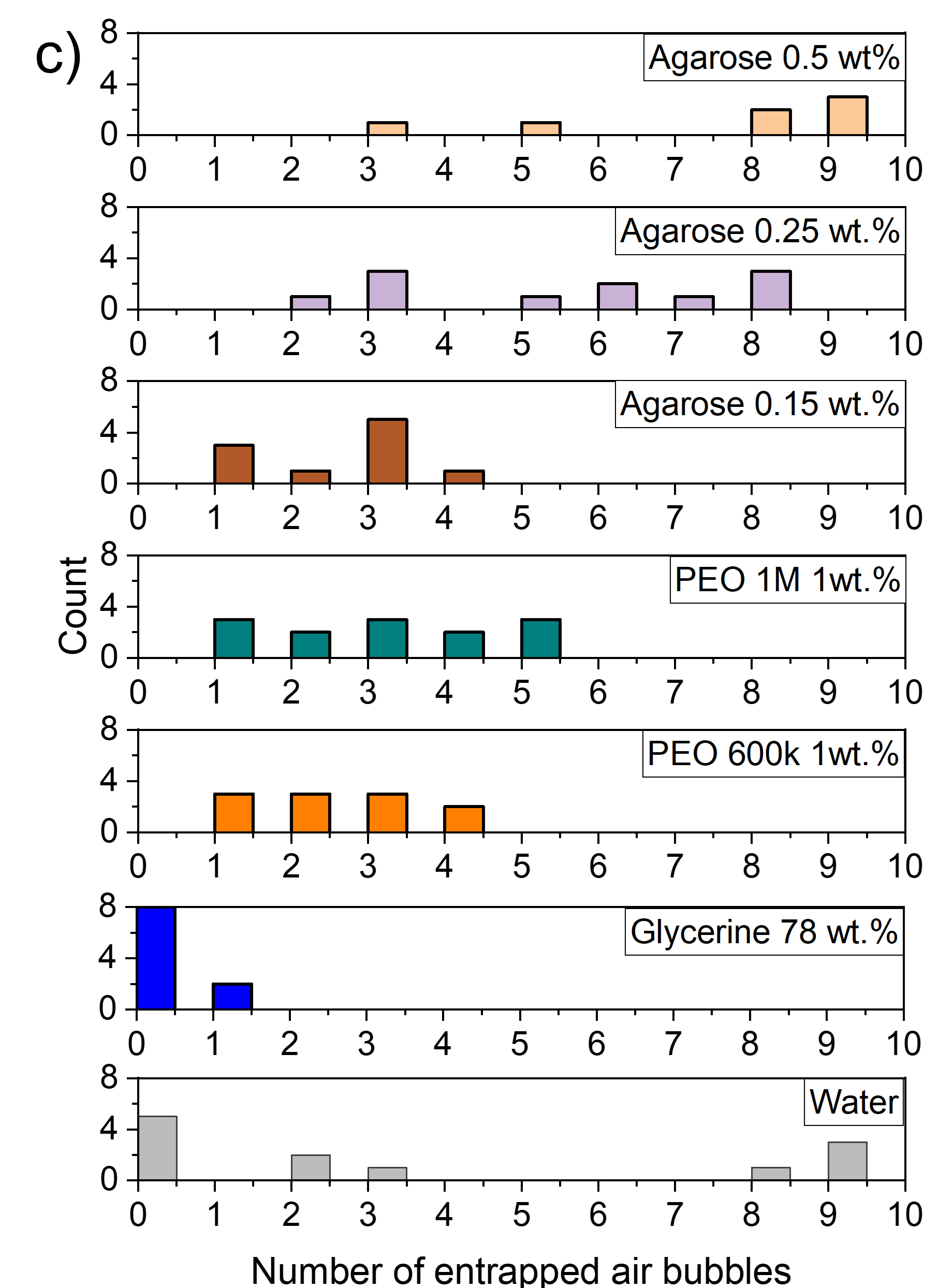}
\caption{Characterisation of bubbles entrained after injection. Total area of entrapped bubbles ($A_{bubbles}$), total injected area ($A_{total}$) and number of entrapped bubbles for each capillary bridge. a) Image analysis used to extract $A_{total}$ and $A_{bubbles}$. Top panel: raw image; middle panel: total injected area ($A_{total}$); bottom panel: area of bubbles $A_{bubbles}$. b) Histogram showing the ratio of the bubble area to total injected area for the liquid and agarose capillary bridges at different impacting conditions. The ratio depends on the closure characteristics. For the `no seal' events no trapped bubbles are observed and in consequence all the injected area is conformed by the liquid jet , e.g., for glyecerine 78 wt.$\%$. In contrast, for surface seal events most the area of the cavity is entrapped as a bubble, thus, the ratio of $A_{bubbles}/A_{total}$ is close to 1. c) Histogram showing the frequency of the number of entrapped droplets for each liquid at different impacting conditions. The number of trapped air bubbles depends on closure characteristics. Several ($\approx$ 4-10) trapped bubbles are the result of the cavity pinching off in different locations. While one or two entrapped bubbles are an indication of the cavity pinching off before the whole volume of the jet is injected into the capillary bridge. In the case of an ideal injection, no bubbles would be trapped and $A_{bubbles}/A_{total} = 0$, i.e., it would look similar to the injections at the glycerine capillary bridge.}
\label{Fig7}
\end{figure}

We also investigate the number and characteristics of entrained bubbles after jet impact on the different capillary bridges. We compare the area of liquid that remains inside the capillary bridge ($A_{total}$) with the area of the entrained bubbles $A_{bubbles}$. An example of the image analysis to obtain $A_{total}$ and $A_{bubbles}$ is given in figure \ref{Fig7}a. The first panel of the figure shows the last frame of the impact of a jet on an agarose 0.15 wt.$\%$ capillary bridge. The second panel shows the binarised image with $A_{total}$, that counts both the area of the injected liquid and $A_{bubbles}$. The last panel shows the binary image showing $A_{bubbles}$ only. These parameters give information about the efficiency of the injection, for example in needle-free applications where air entrainment is undesirable \cite{fleming1999challenging, kalra2012forum}.

In figure \ref{Fig7}b we show the ratio between the total area and the area of the entrapped bubbles $A_{bubbles}/A_{total}$, for the different capillary bridges. A ratio of 0 indicates no bubble entrapment, while a ratio of 1 implies that just bubbles are entrapped no liquid is injected. Figure \ref{Fig7}c shows the number of entrained bubbles for all the experiments. 

For water capillary bridges, $A_{bubbles}/A_{total}$ ranges from 0 to 1. This is the result of the different cavity collapses regimes in water depending on the Weber number (figure \ref{Fig5}a). For example in a `surface seal' the injection results in a bubble area comparable to the maximum cavity area $A_{max}$. In contrast, in the `no seal' regime most of the injected volume comes from the jet and no bubbles are entrapped.

In the case of glycerol capillary bridges, the entrapped air at the end of the injection is negligible, as in all the experiments the cavity collapse was in the `no seal' regime (figure \ref{Fig5}a). Since all the liquid of the jet is injected into the capillary bridge, this is the ideal situation for needle-free injection applications. 

PEO solutions, present the cavity collapse regimes of `deep seal' and the `no seal'. In the `deep seal' regime the cavity collapses after some liquid is already stationary inside the capillary bridge. Consequently, $A_{bubbles}/A_{total}$ did not exceed $0.6$. Interestingly, the distribution of $A_{bubbles}/A_{total}$ and number of entrained bubbles of PEO 1M 1wt.$\%$ is similar to the distribution in agarose 0.15 wt.$\%$. We attribute the similarity to the resemblance in cavity shape and sizes in both of the materials. 

For agarose capillary bridges $A_{bubbles}/A_{total}$ is always close to $\approx 0.5$. This is a consequence that only the `shallow seal' regime was observed for these gels. In addition, the size of the trapped bubbles in agarose is in general smaller than on liquids. This is explained by the cavity collapse at multiple points before the end of the injection. Furthermore, the area of the cavity is smaller for agarose capillary bridges as compared to liquid ones. 


The studies on trapped bubbles are relevant to needle-free applications, specially on agarose gels as they are widely used as skin surrogates \cite{oyarte2019high, sun2017phase}. Therefore, the consideration of trapped bubbles and how to minimise their occurrence and volume need to be addressed, when injecting onto real skin. We can conclude that preventing the cavity collapse at the surface of the injection is crucial to avoid entrapping bubbles larger in volume than the injected liquid. However, as skin resembles more agarose than water, and we did not observe surface seal for agarose, we deem the `surface seal' regime unlikely for skin. Furthermore, by promoting the `no seal' regime, all the liquid of the jet would be injected without air entrapment.   

\subsection{{Surface wettability effects}}

\begin{figure}
\centering
\includegraphics[width=15cm]{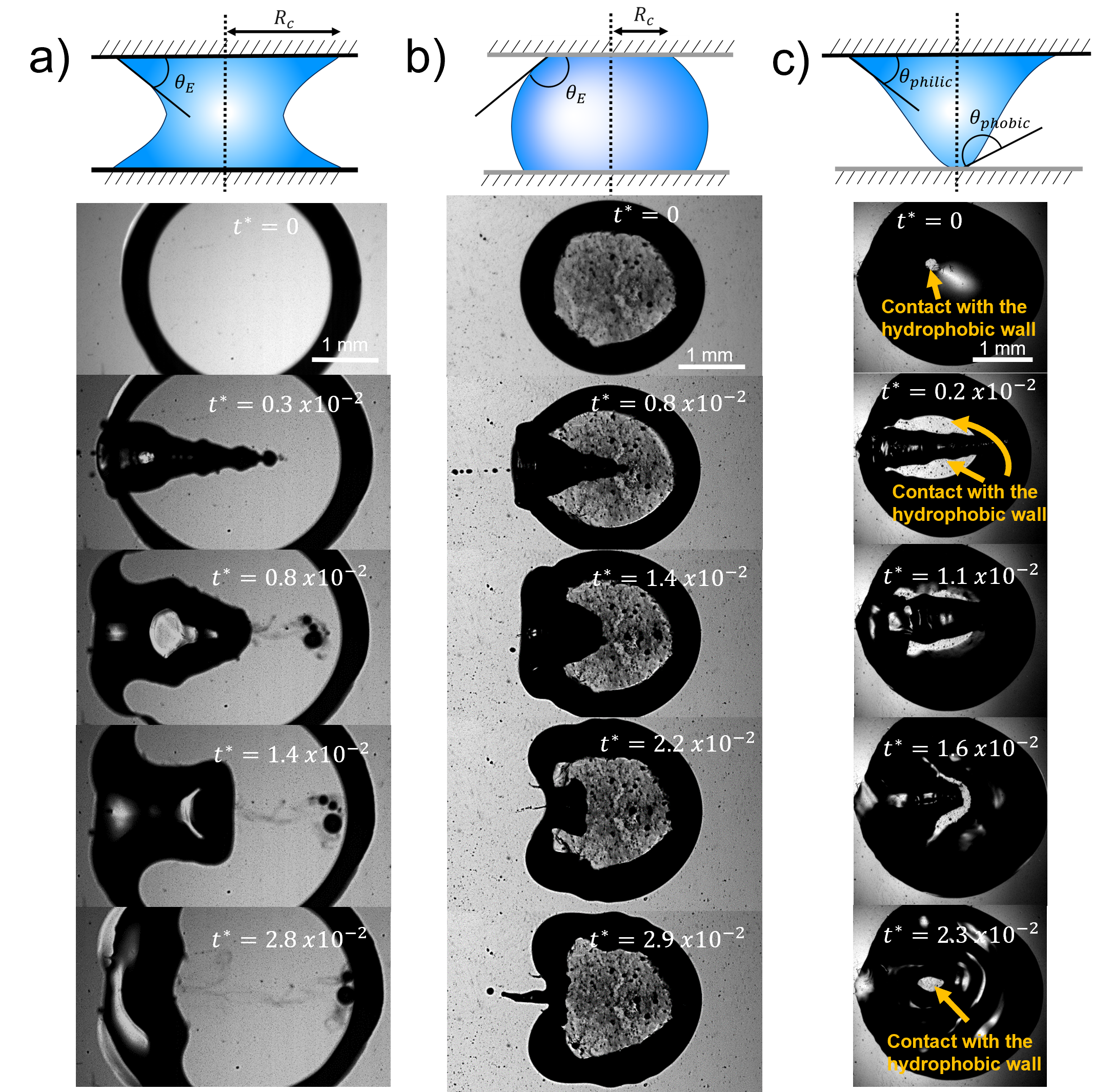}
\caption{Schematics of the capillary bridge geometry and snapshots sequences of the retraction of the cavity generated after the impact of a jet on a capillary bridge. a) Capillary bridge between hydrophilic surfaces ($We= 352$; Movie 6 in the supplementary material). In this case the capillary bridge is concave. b) Capillary bridge between hydrophobic surfaces ($We= 325$; Movie 7 in the supplementary material). In this case the interface of liquid bridge is convex. c) Capillary bridge with a hydrophilic surface on the top and a hydrophobic on the bottom: hydrophilic-hydrophobic ($We = 320$; Movie 8 in the supplementary material). Here, the capillary bridge changes from curvature and adopts a \textit{conical} shape. After the impact, for the hydrophobic surfaces, the contact line moves and a Worthington jet is observed. In contrast for the hydrophilic surfaces, as well as for the hydrophilic-hydrophobic surfaces the contact line remains pinned and no Worthington jet is observed.}
\label{Fig8}
\end{figure}

\begin{figure}
\centering
\includegraphics[width=12cm]{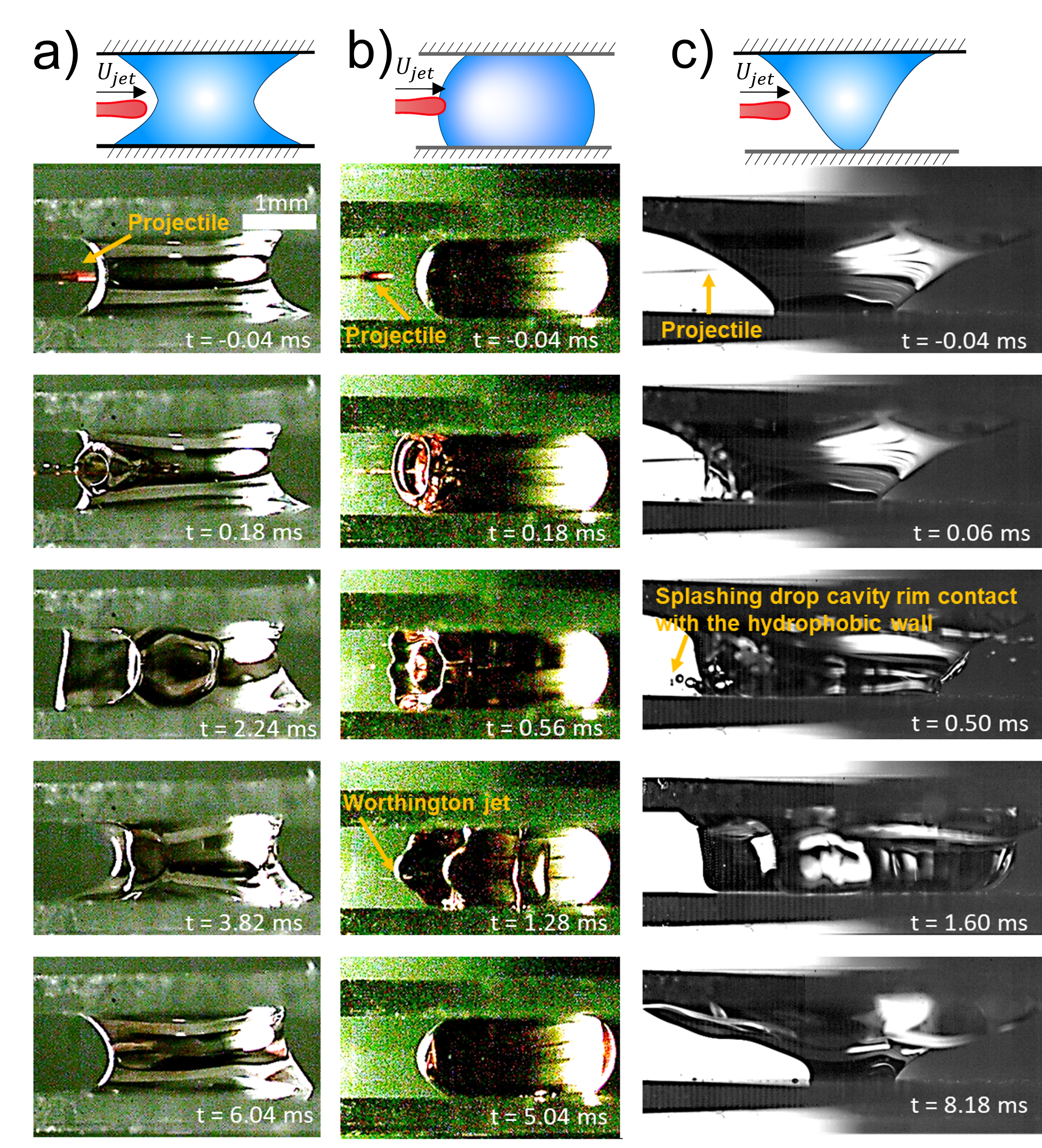}
\caption{Snapshots sequences of the front view of the impact of a jet onto a capillary bridge. a) Capillary bridge between hydrophilic surfaces ($We= 468$). At t = 0.18 a cavity is created with its respective rim. While the cavity expands, the rim touches the walls and remains pinned (Movie 9 in the supplementary material). b) Capillary bridge between hydrophobic surfaces ($We= 408$). In this case, when the rim touches the walls, it bounces back. Due to this bouncing the contact line moves. At time t = 1.28 ms droplets coming from the Worthington jet can be observed (Movie 10 in the supplementary material). c) Capillary bridge with a hydrophilic surface on the top and a hydrophobic on the bottom: hydrophilic-hydrophobic ($We = 1563$). Here, when the rim grows and contacts the walls, the liquid pins in the hydrophilic wall, while it splashes in the hydrophobic wall. At time t = 1.60 ms most of the liquid detaches from the hydrophobic wall, and at time t = 8.18 ms, the liquid bridge is almost at the equilibrium position (Movie 11 in the supplementary material).}
\label{Fig9}
\end{figure}

In this section we change the wettability of the walls where the capillary bridge is contained and asses its effect on the impact process. Here, we applied Glaco coating on the glass walls (originally hydrophilic) to render them hydrophobic. Figures \ref{Fig8}, and \ref{Fig9} present snapshots of the experiments in the perpendicular view and at an angle. In figure \ref{Fig8} we show the retraction dynamics of the cavity formed after a jet impact on a capillary bridge that is between either hydrophobic walls (figure \ref{Fig8} a), hydrophilic walls (figure \ref{Fig8} b) or a mix hydrophilic on the top and hydrophobic on the bottom (hydrophilic-hydrophobic, figure \ref{Fig8} c). The cavity retraction ends with a Worthington jet on the hydrophobic walls. In contrast, no Worthington jet is observed during the retraction on hydrophilic and hydrophilic-hydrophobic walls. Moreover, we observed that the change of wettability is negligible during cavity expansion, as expected from an inertia dominated process (see Section III.A). The only difference being the fate of the rim of the cavity when it touches the wall, where it can stay pinned or bounce (see figure \ref{Fig9} and supplementary Movies 9-11). 
First, we study the hydrophilic-hydrophobic capillary bridge to understand the difference in the contact line dynamics between the two surfaces. In this configuration, before impact, the liquid is touching only a \textit{point} on the hydrophobic surface, while wetting an area of 6.3 mm$^2$ on the hydrophilic wall (see figure \ref{Fig8}c at $t^*$ = 0 and first panel of figure \ref{Fig9}c). During cavity expansion the liquid of the capillary bridge moves from the impact point towards the hydrophobic wall and wets it (see figure \ref{Fig8}c at $t^*$ = $0.2 \times 10^{-2}$ ). If the impact force is enough the rim will splash after contacting the hydrophobic wall as seen in figure \ref{Fig9} and supplementary video 11. When the cavity recedes and the capillary bridge reaches equilibrium the liquid dewetts the wall and its state is similar than before the impact. In contrast the contact line stays pinned in the hydrophilic wall for the whole process.


Figure \ref{Fig8} shows the geometry of the capillary bridge for hydrophobic, hydrophilic and hydrophilic-hydrophobic walls. We note that the liquid bridge is concave for hydrophilic walls and convex for hydrophobic walls. For the hydrophilic-hydrophobic walls the capillary bridge changes from curvature near the walls and adopts a \textit{conical} shape. This curvature changes are the result from the equilibrium contact angle of the surfaces and surface energy minimisation. The difference in curvature introduces variations in the Laplace pressure $\Delta P$ between the different capillary bridges. The Laplace pressure of a capillary bridge of height $H$ with a contact angle $\theta_E$ is \cite{wang2019effect},

\begin{equation}
    \Delta P = \gamma_{cb} \left(\frac{2}{D_{cb}} - \frac{cos\theta_E}{H/2}\right) \approx \frac{-2\gamma_{cb} cos\theta_E}{H}.
    \label{Laplace}
\end{equation}

From equation \ref{Laplace}, we observe that the Laplace pressure is positive for a convex bridge and negative for a concave bridge as for $cos(\theta_E) > 0$ for $\theta_E < \pi/2$ and $cos(\theta_E) < 0$ for $\theta_E > \pi/2$ \cite{wang2019effect}. 

The vertical adhesive force $F_a$ of the liquid acting on the walls $F_a$, is given by the Laplace pressure and the axial component of the surface tension acting along the contact line \cite{orr1975pendular,qian2006scaling}. The contact line force $F_{cl}$ is given by the sine of $\theta_E$, the perimeter of the contact area $2\pi R_c$ and the surface tension of the capillary bridge $\gamma_{cb}$. Thus, the vertical adhesive force can be written as  \cite{chen2013modeling},

\begin{equation}
    F_a = 2\pi\gamma_{cb} R_c sin\theta_E - \pi R_c^2 \Delta P,
    \label{Adhesion}
\end{equation}

where $R_c$ is the contact radius of the capillary bridge. The contact radius is also influenced by the surface wettability and for the same liquid volume $R_c$ is larger for a hydrophilic surface than for a hydrophobic one. A typical contact radius for a 2 $\mu$L droplet on glass surfaces is $\approx$ 2.5 mm, while for a water droplet on the Glaco coated surfaces is $\approx$ 1.9 mm. For positive $F_a$, the liquid is attracted by the walls, while negative values of $F_a$ suggest repulsion from the walls \cite{chen2013modeling}. In our experiments glass has a contact angle $\theta_E = 23$ degrees with water, and the Glaco coated glass has a contact angle $\theta_E = 160$ degrees with water. Therefore, for our experiments a typical adhesion force on a capillary bridge between glass walls is $\approx$ 3 mN and for Glaco sprayed glass is $\approx$ -1.2 mN. 
During a fluid-fluid displacement system there is significant energy dissipation from the contact line \cite{primkulov2020characterizing}. In particular, when a system is confined, the ratio of the interfacial area to bulk volume increases, rendering the contact line dissipation more prominent. Contact angle hysteresis, defined as difference between the advancing and receding contact angles ($\theta_a$ and $\theta_r$ respectively), is responsible of the adhesion of drops to inclined surfaces and determines the contact line dissipation $\Phi$, which can be written as \cite{good1992contact,bonn2009wetting},

\begin{equation}
    \Phi_{cl} = 2\pi R_c \gamma_{cb} (cos\theta_a - cos\theta_r).
    \label{Dissipation}
\end{equation}

Hydrophilic surfaces generally have a larger contact angle hysteresis than hydrophobic surfaces \cite{quetzeri2019role}. More specifically, as for Glaco sprayed surfaces water rests on a thin air layer, liquids are very mobile, and there is an almost negligible contact angle hysteresis \cite{quetzeri2019role}. In contrast, surface heterogeneity causes hysteresis in water, which we measured to be $\approx$ 18 degrees. From equation \ref{Dissipation} we get that the contact line dissipation in the glass surfaces is $\approx$ 0.08 mJ, which is comparable to the kinetic energy of the jet $E_{k_{jet}} \approx (\pi/8) \rho_{0}U_{0}^2 D_{0}^2 L_{0} \approx$ 0.12 mJ, where $L_0$ is the length of the jet. Consequently, we expect that the jet does not have enough energy to make the contact line dewett the walls. 

In summary, the Worthington jet is not energetically favourable for the capillary bridge confined between hydrophilic surfaces. The contact line dissipation and adhesion forces in this configuration are comparable to $E_{k_{jet}}$. In contrast, for hydrophobic walls, the adhesion force and contact line dissipation are negligible, allowing contact line movement, thus a Worthington jet becomes energetically favourable.

\section{Conclusions}

We  presented experimental results on the impact of a microfluidic onto liquid and agarose capillary bridges. By using high speed imaging and image analysis we extracted the cavity profiles across the whole duration of the experiment. 

We modelled the cavity expansion and shape based on the comparison between the Young–Laplace and the dynamic pressures of the cavity made by the penetrating jet. We compared the model of the cavity shapes (summarised by equation \ref{CavityProfile}) with experiments, finding good agreement between the experiments and model for the cavity shape and evolution on the water capillary bridge. However, for the impact on the glycerol solution we observed that the radial cavity expansion is overestimated. We attribute this mismatch to viscous losses, which are neglected in equation \ref{CavityProfile}. By using equations 5 and 6 and the simile between a jet and a droplet train, we arrived to an expression that predicts the cavity profile that depends on the Reynolds number.

Furthermore, we found that the Weber number threshold for traversing a capillary bridge is larger than for a pendant droplet. We attributed this to the confinement and dissipation of energy along the walls. Equation \ref{GammaDyn} predicts the threshold with $10\%$ accuracy for Newtonian liquids and the agarose gel 0.15 wt.$\%$. In contrast, for agarose gels 0.5 wt.$\%$, the predicted threshold is smaller by one order of magnitude. We attribute this deviation to an overestimation of the surface tension predicted by equation \ref{SurfaceTensionAgarose}. This indicates, that as the shear modulus of the gels are increased, they can no longer be treated as liquids. Hence, for needle-free applications, more complex models for predicting the depth of injection in skin need to be implemented. This finding should be useful for the bioenginering community that uses hydrogels as skin surrogate.

Additionally, we studied the cavity collapse, the number of trapped bubbles and the ratio between the area of the bubbles and the injected liquid. For the range of Weber numbers studied here, the water cavities presented a varied type of collapse. Consequently, the number of trapped bubbles varied from 0 to 10 and $A_{bubbles}/A_{total}$ was in the range from 0 to 1. The aqueous glycerol solution had a single collapse mode and no bubbles were trapped. Finally, for the agarose gels, the cavity was observed to collapse in multiple points trapping several bubbles that amounted to an area approximately half of the injected area. Accordingly, for needle-free injections, the ideal scenario is the no seal regime. In contrast, the surface seal regime, should be avoided. However, as skin is viscoelastic and has a similar storage modulus to agarose, we do not expect this scenario to happen in the range of velocities leading to injection. 

Lastly, we assessed how the wettability of the walls confining the capillary bridge influenced the cavity collapse. We argue that the Worthington jet is suppressed for the hydrophilic walls due to a larger adhesion force (equation \ref{Adhesion}) and the contact line dissipation of energy (equation \ref{Dissipation}), as compared to the hydrophobic walls.

Our results bridge the gap between the knowledge on the comparisons between the impact on liquids and soft solids, such as agarose gels. It also, gives insight on the amount of liquid and trapped bubbles generated after the impact of a microfluidic jet onto a confined target. A better control and understanding of how the cavity collapses presents an important advance towards the control of needle-free injections. 

\section*{Conflict of interests}
There are no conflicts to declare.

\section*{Acknowledgements}
This research was funded by the European Research Council (ERC) under the European Union Horizon 2020 Research and Innovation Programme (grant agreement no. 851630).

\bibliography{apssamp}

\end{document}